\documentclass[twocolumn,prb,showpacs,preprintnumbers,amsmath,amssymb,superscriptaddress]{revtex4-1}
\usepackage{graphicx}
\usepackage{setspace}

\usepackage{color}
\usepackage{ulem} 
\graphicspath{{./figures/}}

\usepackage{amsmath,amsthm,amssymb}
\usepackage{mathrsfs}

\usepackage[
colorlinks=true, citecolor=blue, urlcolor=blue, linkcolor=blue,
setpagesize=false, bookmarks=false, breaklinks=true
]{hyperref}

\newcommand{\hc}{\hat{c}}
\newcommand{\hd}{\hat{d}}

\newcommand{\heta}{\hat{\eta}}
\newcommand{\hs}{\hat{s}}
\newcommand{\hH}{\hat{H}}
\newcommand{\hA}{\hat{A}}
\newcommand{\hB}{\hat{B}}
\newcommand{\hn}{\hat{n}}
\newcommand{\hN}{\hat{N}}

\newcommand{\hS}{\hat{S}}
\newcommand{\hK}{\hat{K}}
\newcommand{\hrho}{\hat{\rho}}

\newcommand{\eqq}[1]{\begin{align} #1 \end{align}}

\begin{document}

\title{Exploring nonequilibrium phases of photo-doped Mott insulators with Generalized Gibbs ensembles}

\author{Yuta Murakami$^\dagger$}
\affiliation{Department of Physics, Tokyo Institute of Technology, Meguro, Tokyo 152-8551, Japan\\
yuta.murakami@phys.titech.ac.jp}
\author{Shintaro Takayoshi}
\affiliation{Department of Physics, Konan University, Kobe 658-8501, Japan}
\author{Tatsuya Kaneko}
\affiliation{Department of Physics, Columbia University, New York, New York 10027, USA}
\author{Zhiyuan Sun}
\affiliation{Department of Physics, Columbia University, New York, New York 10027, USA}
\author{Denis Gole\v{z}}
\affiliation{Jozef Stefan Institute, Jamova 39, SI-1000 Ljubljana, Slovenia}
\affiliation{Faculty of Mathematics and Physics, University of Ljubljana, Jadranska 19, 1000 Ljubljana, Slovenia }
\author{Andrew J. Millis}
\affiliation{Department of Physics, Columbia University, New York, New York 10027, USA}
\affiliation{Center for Computational Quantum Physics, Flatiron Institute, New York, New York 10010, USA}
\author{Philipp Werner}
\affiliation{Department of Physics, University of Fribourg, 1700 Fribourg, Switzerland}

\begin{abstract}
Photo-excited strongly correlated systems can exhibit intriguing non-thermal phases, but the theoretical investigation of them poses significant challenges. In this work, we introduce a generalized Gibbs ensemble type description for long-lived photo-doped states in Mott insulators. This framework enables systematic studies of photo-induced phases based on equilibrium methods, as demonstrated here for the one-dimensional extended Hubbard model.
We determine the nonequilibrium phase diagram, which features $\eta$-pairing and charge density wave phases in a wide doping range, and reveal physical properties of these phases. We show that the peculiar kinematics of photo-doped carriers, and the interaction between them, play an essential role in the formation of the non-thermal phases, and we clarify the differences between photo-doped Mott insulators, chemically-doped Mott insulators and photo-doped semiconductors. Our results demonstrate a new path for the systematic exploration of nonequilibrium strongly correlated systems and show that photo-doped Mott insulators host different phases than conventional semiconductors.
\end{abstract}

\maketitle
\noindent{\bf Introduction}\\
Nonequilibrium control of quantum materials is an intriguing prospect with potentially important technological applications ~\cite{Yonemitsu2008,Giannetti2016review,Basov2017review,Cavalleri2018review,Oka2019review,Sentef2021nonthermal}.
Experiments with various materials and excitation conditions have reported phenomena not observable in equilibrium including nonthermal ordered phases such as superconducting (SC)-like phases~\cite{Cavalleri2011,Mitrano2016,Cavalleri2018review,Suzuki2019}, charge density waves (CDW)~\cite{Mihailovic2014,Okamoto2014PRL,Kogar2020} and excitonic condensation~\cite{Murotani2019PRL}.
Among various nonequilibrium protocols, photo-doping is a basic and important one, in which a radiation pulse creates electron- and hole-like charge carriers with a long lifetime on the electronic timescale.
Due to the nonthermal distribution of the carriers and the cooperative interplay between them, peculiar nonequilibrium phases can be induced.

The theory of photo-doping has been extensively discussed for semiconductors,
where a rich phase diagram including electron-hole plasmas and exciton gases~\cite{Mott1961,Haug1982,Zimmermann1985,haug_quantum_1990,Keldysh1986review,Asano2014JPSJ} as well as exciton condensation~\cite{Keldish1965,Rice1967EI,Perfetto2019PRM} is found.
The physical picture is that, after electrons and holes are created, they rapidly relax within the conduction and valence bands, while their recombination occurs on a much longer timescale. 
Thus, at the single-particle level the numbers of electrons and holes are separately conserved, so that in the intermediate time regime one has a pseudoequilibrium state that can be described by an effective equilibrium theory with separate chemical potentials for the electrons and holes~\cite{haug_quantum_1990,Keldysh1986review,Asano2014JPSJ}.

A situation of great current interest is the photodoping of Mott insulators. 
In these systems, exotic equilibrium states such as unconventional SC phases emerge upon chemical doping~\cite{Tokura_RMP},
while photo-doping creates novel pseudoparticle excitations not easily represented in a single particle picture, e.g. doublons and holons in the single-band case. 
When the Mott gap is large enough, these excitations are long-lived due to the lack of efficient recombination channels~\cite{Strohmaier2010PRL,Zala2013PRL,Mitrano2014,Sensarma2010PRB,Martin2011PRB,Zala2014PRB}.
Therefore, as in semiconductors, a fast intraband relaxation results in a long-lived quasi-steady state.
Previous studies based on short-time simulations indicated the emergence of enhanced CDW~\cite{Tohyama2012PRL} or SC correlations~\cite{Wang2018PRL,Nikolaj2019JPSJ,Kaneko2019PRL,Kaneko2020PRR,Ejima2020PRR}, as well as novel spin-orbital orders~\cite{Jiajun2018Natcom}.
However, unlike in the semiconductor cases, the long-time behavior of photo-doped Mott insulators is not well understood, due to the lack of powerful theoretical frameworks.

Steady state formalisms in which explicit heat/particle baths or other dissipative mechanisms are attached to the system have been recently applied~\cite{Jiajun2020PRB,Jiajun2021PRB,Jaksch2019PRL}.
These formalisms, however, require attention to the influence of the baths and dissipations, and the use of explicitly nonequilibrium methods.
An alternative approach is a pseudoequilibrium description as in conventional semiconductors. 
Crucial differences from semiconductors are that the approximately conserved entities are pseudoparticles (local many-body states), with which the Hamiltonian is not initially specified, 
and that the approximate conservation of their number is not manifest in the Hamiltonian but arises from kinematic constraints. 
Previous works~\cite{Takahashi2002PRB,Takahashi2005PRB,Rosch2008PRL,Ishihara2011PRL,Jiajun2020PRB} introduced the idea of using the Schrieffer-Wolff (SW) transformation 
to reformulate the problem in a way that explicitly references the approximately conserved quantities and isolates the terms that eventually lead to full equilibration.
However, the applications of  such effective descriptions have been so far limited to  small clusters and weak~\cite{Takahashi2002PRB,Takahashi2005PRB,Ishihara2011PRL} or extreme excitation conditions~\cite{Rosch2008PRL}.

In this work, we introduce a generalized Gibbs ensemble (GGE) type description for the effective model obtained from a SW transformation by incorporating different {\it chemical potentials for pseudoparticles}, i.e. for the many-body local states.
This effective equilibrium description allows us to systematically scan the nonequilibrium states in photo-doped Mott insulators for extended systems using established equilibrium methods.
We use this approach to identify and study emerging phases in the photo-doped one-dimensional extended Hubbard model.
We determine the nonequilibrium phase diagram, where $\eta$-pairing~\cite{Yang1989PRL,Rosch2008PRL,Kaneko2019PRL,Ejima2020PRR,Jiajun2020PRB} and CDW phases appear in a wide doping range, and reveal the corresponding spectral features.
The CDW phase is strongly favored in photo-doped systems, compared to the chemically-doped ones, and it is characterized by unbound doublons and holons in contrast to photo-doped semiconductors where electron-hole binding is an important effect. 
We show that the kinematics of photo-doped doublon/holon carriers, which is qualitatively different from the electron/hole dynamics in conventional semiconductors, plays an essential role and leads to development of CDW and $\eta$-pairing correlations described by squeezed systems without singly occupied sites. \\

\noindent{\bf Results}\\
{\bf GGE description for photo-doped Mott insulators.} 
The generic formulation of the GGE description is  given in the Supplemental Material (SM), but here we explain the procedure focussing on the extended Hubbard model,
whose Hamiltonian is 
\eqq{
\hat{H} =-t_{\rm hop}\sum_{\langle i,j\rangle,\sigma} (c^\dagger_{i,\sigma}c_{j,\sigma} +h.c.) + \hH_U + \hH_V,
}
with $\hH_U=U \sum_i (\hn_{i\uparrow}-\frac{1}{2}) ( \hn_{i\downarrow}-\frac{1}{2}) $ the on-site and $ \hH_V \nonumber=V \sum_{\langle i,j\rangle } (\hn_i-1) (\hn_{j}-1)$ the nearest-neighbor interaction. $\hc^\dagger_{i\sigma}$ is the creation operator of a fermion with spin $\sigma$ at site $i$, 
$\hn_{i\sigma} =\hc^\dagger_{i\sigma} \hc_{i\sigma}$ the spin-density at site $i$, $\hn_{i} = \hn_{i\uparrow} + \hn_{i\downarrow}$, and $\langle i,j\rangle$ denotes pairs of nearest-neighbor sites. 
$v$ is the hopping parameter.  For large $U$, the half-filled equilibrium system is Mott insulating. 

Photo-doping the Mott insulator creates doublons (doubly occupied states) and holons (empty states).
When the Mott gap is large, the number of these excited local states is approximately conserved for kinematic reasons~\cite{Strohmaier2010PRL,Zala2013PRL,Mitrano2014,Sensarma2010PRB,Martin2011PRB,Zala2014PRB}.
However, the original Hamiltonian explicitly contains recombination terms, which generate virtual processes that affect the physics even when recombination is kinematically suppressed.
In order to explicitly remove the recombination terms, while effectively taking account of effects of virtual recombination processes, we perform the SW transformation~\cite{SW_original, Jiajun2020PRB, Murakami2021PRB}.
Here, we assume $U\gg V,t_{\rm hop}$. 
The effective Hamiltonian up to $\mathcal{O}(t_{\rm hop}^2/U)$ is 
\eqq{
 \hH_{\rm eff}  =&   \hH_U + \hH_{\rm kin,holon} +  \hH_{\rm kin,doub} + \hH_V \nonumber\\
 &                          +  \hH_{\rm spin,ex} +  \hH_{\rm dh,ex} + \hH^{(2)}_{U,\rm{shift}} + \hH_{\rm 3-site}, \label{eq:Heff}
}
where $ \hH_{\rm kin,holon}$ and $ \hH_{\rm kin,doub}$ describe the hopping of holons and doublons of $\mathcal{O}(t_{\rm hop})$, respectively.
The remaining terms are of $\mathcal{O}(t_{\rm hop}^2/U)$.
$ \hH_{\rm spin,ex}$ is the spin exchange term, $\hH_{\rm dh,ex}$ is the doublon-holon exchange term and $\hH^{(2)}_{U,\rm{shift}}$ describes the shift of the local interaction.
$\hH_{\rm 3-site}$ represents three-site terms such as correlated doublon hoppings, see Method.
$\hH_{\rm dh,ex}$ sets the correlations between the neighboring doublons and holons as $\hH_{\rm spin,ex}$ sets spin correlations between singlons (singly occupied states).
In this sense, the above model is a natural extension of the $t$-$J$ model~\cite{Tokura_RMP}, which is obtained by assuming that either holons or doublons are added (chemical doping) and ignoring $\hH_{\rm 3-site}$.

Due to intraband scatterings and environmental couplings (e.g. phonons), intraband relaxation occurs and the system reaches a steady state.
Since the numbers of doublons and holons are conserved in the effective model, the steady state can be described by introducing separate {\it chemical potentials} for them. 
The corresponding number operators are $\hat{N}_{\rm holon} = \sum_i \hn_{i,h}$ with $\hn_{i,h} =(1-\hn_{i\uparrow})(1-\hn_{i\downarrow})$ and $\hat{N}_{\rm doub} = \sum_i \hn_{i,d}$ with $\hn_{i,d}=\hn_{i\uparrow} \hn_{i\downarrow}$,
respectively.  With these, the grand-canonical Hamiltonian $\hK_{\rm eff}$ can be written as $ \hH_{\rm eff}-\mu_{\rm holon} \hat{N}_{\rm holon} -\mu_{\rm doub} \hat{N}_{\rm doub}$, or 
\eqq{
\hK_{\rm eff}  = \hH_{\rm eff}  - \mu_U \sum_i \hn_{i\uparrow} \hn_{i\downarrow} -\mu \sum_i  \hn_{i},
}
where $\mu_U = \mu_{\rm doub} + \mu_{\rm holon}$ and $\mu= -\mu_{\rm holon}$.
Thus, the local interaction is modified from $U$ by the photo-doping ($U-\mu_U$), i.e. the energy difference between the doublons and holons is effectively reduced, analogously to the effective shift of the band splitting in photo-doped semiconductors \cite{Perfetto2019PRM,Murakami2020PRB}.
The properties of the nonequilibrium steady states may then be described by the density matrix $\hrho_{\rm eff} = \exp(-\beta_{\rm eff} \hat{K}_{\rm eff})$ with an effective temperature $T_{\rm eff}=1/\beta_{\rm eff}$~\cite{haug_quantum_1990,Asano2011PRL,Perfetto2019PRM}, which is a sort of GGE~\cite{Vidmer2016,Zala2017Natcom}.
This is essentially an equilibrium problem that can be studied with established equilibrium techniques.
Response functions $-i\langle [\hat{A}(t),\hat{B}(0)]_{\rm \pm} \rangle$ of the nonequilibrium states can also be computed within this framework, see SM.
Our basic assumption that the nonequilibrium states can be characterized by a few parameters, such as the doublon number and effective temperature, is supported by a recent study~\cite{Jiajun2020PRB}, which demonstrated a good agreement between time-evolving states and nonequilibrium steady states weakly coupled to  thermal baths.

\noindent {\bf Nonequilibrium phases diagram of photo-doped extended Hubbard model.} 
We apply the above framework to the half-filled  one-dimensional extended Hubbard model using infinite time-evolving block decimation (iTEBD))~\cite{Vidal2003PRL} and exact diagonalization (ED)~\cite{Dagotto1994RMP}.
As in the case of the $t$-$J$ model, the effect of $\hH_{\rm 3-site}$ is not essential. We confirm this for the photo-doped situation using ED in the SM.
Thus, in the following, we focus on the effective model $\hH_{\rm eff2}$, which ignores $\hH_{\rm 3-site}$.
We consider cold systems ($T_{\rm eff}=0$) to clarify the possible emergence of nonequilibrium ordered phases. 
Such a situation may be achieved by energy dissipation to the environment~\cite{Eckstein2013,Sentef2013, Murakami2020PRB} or entropy reshuffling~\cite{Werner2019PRB,Werner2019Natcom}.
In the following, we use $t_{\rm hop}$ as the energy unit, and fix $U=10$, i.e. $J_{\rm ex}\equiv\frac{4t_{\rm hop}^2}{U}$ = $0.4$.
We examine the charge correlations $\chi_{c}(r) \equiv \frac{1}{N}\sum_i \langle (\hn_{i+r}-n_{\rm av}) (\hn_i-n_{\rm av}) \rangle$,
 spin correlations $\chi_{s}(r) \equiv  \frac{1}{N}\sum_i \langle \hS^z_{i+r} \hS^z_i \rangle $, and SC correlations $\chi_{sc}(r)\equiv \frac{1}{N}\sum_i \langle \hat{\Delta}^\dagger_{i+r} \hat{\Delta}_i \rangle$.
 Here, $N$ is the system size, $n_{\rm av}= \frac{1}{N}\sum_i \langle \hn_i\rangle$, $\hS_i^z = \frac{1}{2}(\hn_{i,\uparrow}-\hn_{j,\downarrow})$, and $\hat{\Delta}_i=\hc_{i\uparrow}\hc_{i\downarrow}$.
$\eta$-pairing is characterized by staggered SC correlations.
Note that $\hH$, $\hH_{\rm eff}$, and $\hH_{\rm eff2}$ are SU$_c(2)$ symmetric with respect to the $\eta$-operators for $V=0$~\cite{Yang1989PRL,Essler2005}.
Due to this symmetry, a homogeneous state with long-range $\eta$-SC correlations is on the verge of phase separation, which we avoid by considering non-zero $V$~\cite{Rosch2008PRL}, see SM.

 \begin{figure}[t]
  \centering
    \hspace{-0.cm}
    \vspace{0.0cm}
\includegraphics[width=75mm]{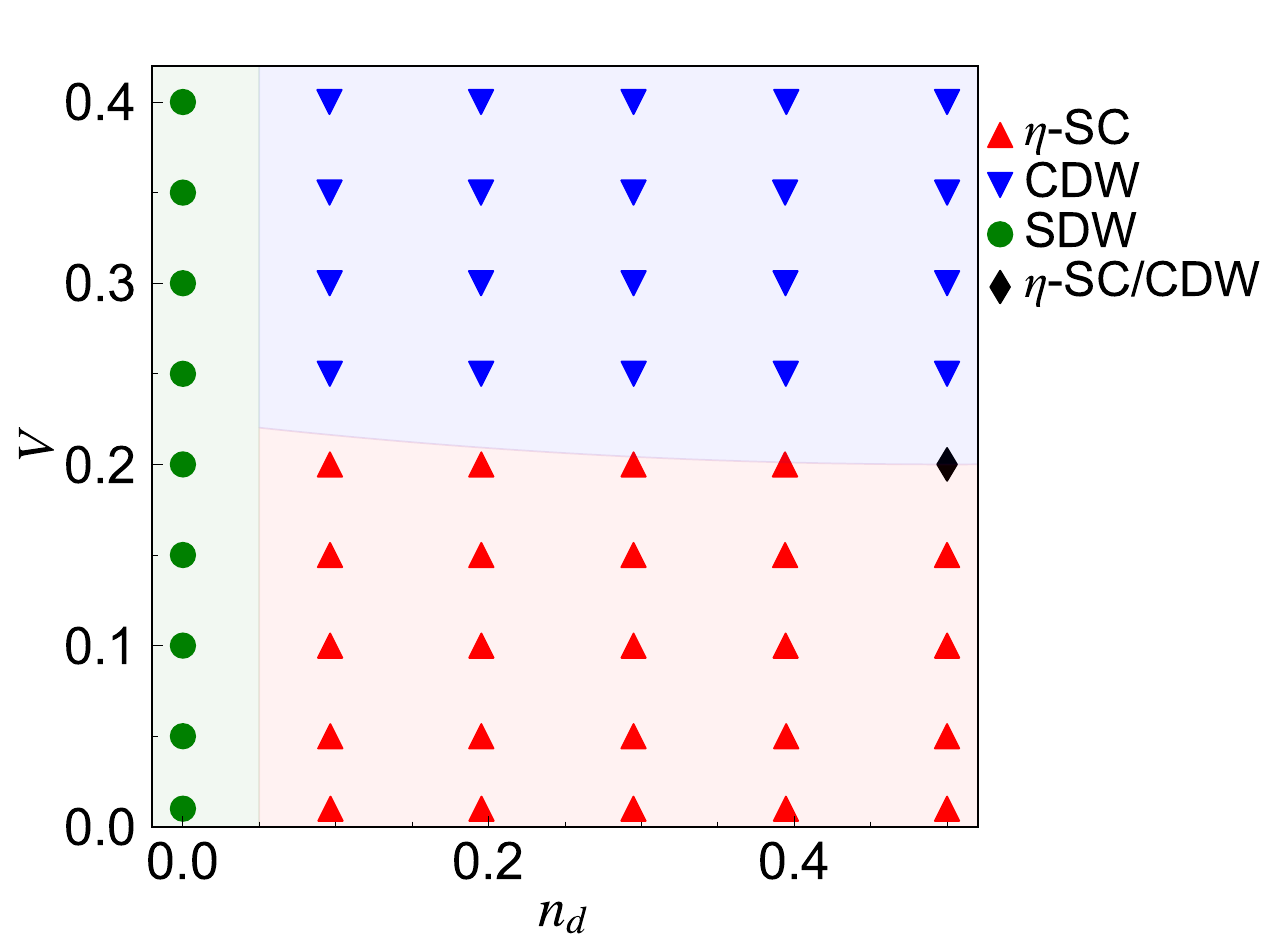} 
  \caption{Phase diagram of the half-filled photo-doped Mott insulator described by $\hH_{\rm eff2}$ in the plane of the doublon density ($n_d= \frac{1}{N}\sum_i \langle \hn_{i,d}\rangle$) and the nonlocal interaction $V$, 
  for $U=10$. Phases are categorized by the dominant correlation evaluated from iTEBD, i.e. the correlation with the smallest critical exponent $a$. 
  The phase boundary is only schematic and a guide to the eye.
  The critical exponent is extracted by fitting the correlation functions ($\chi(r)$) with $C_1/r^2 + C_2\cos(q r)/r^a$, where $q=2n_d\pi$, $q=(1-2n_d)\pi$ and $q=\pi$ for charge, spin and SC correlations, respectively.
  We use $r\in[6,30]$ for the fitting range.
  }
  \label{fig:iTEBD_phase}
\end{figure}

 \begin{figure}[t]
  \centering
    \hspace{-0.cm}
    \vspace{0.0cm}
\includegraphics[width=64mm]{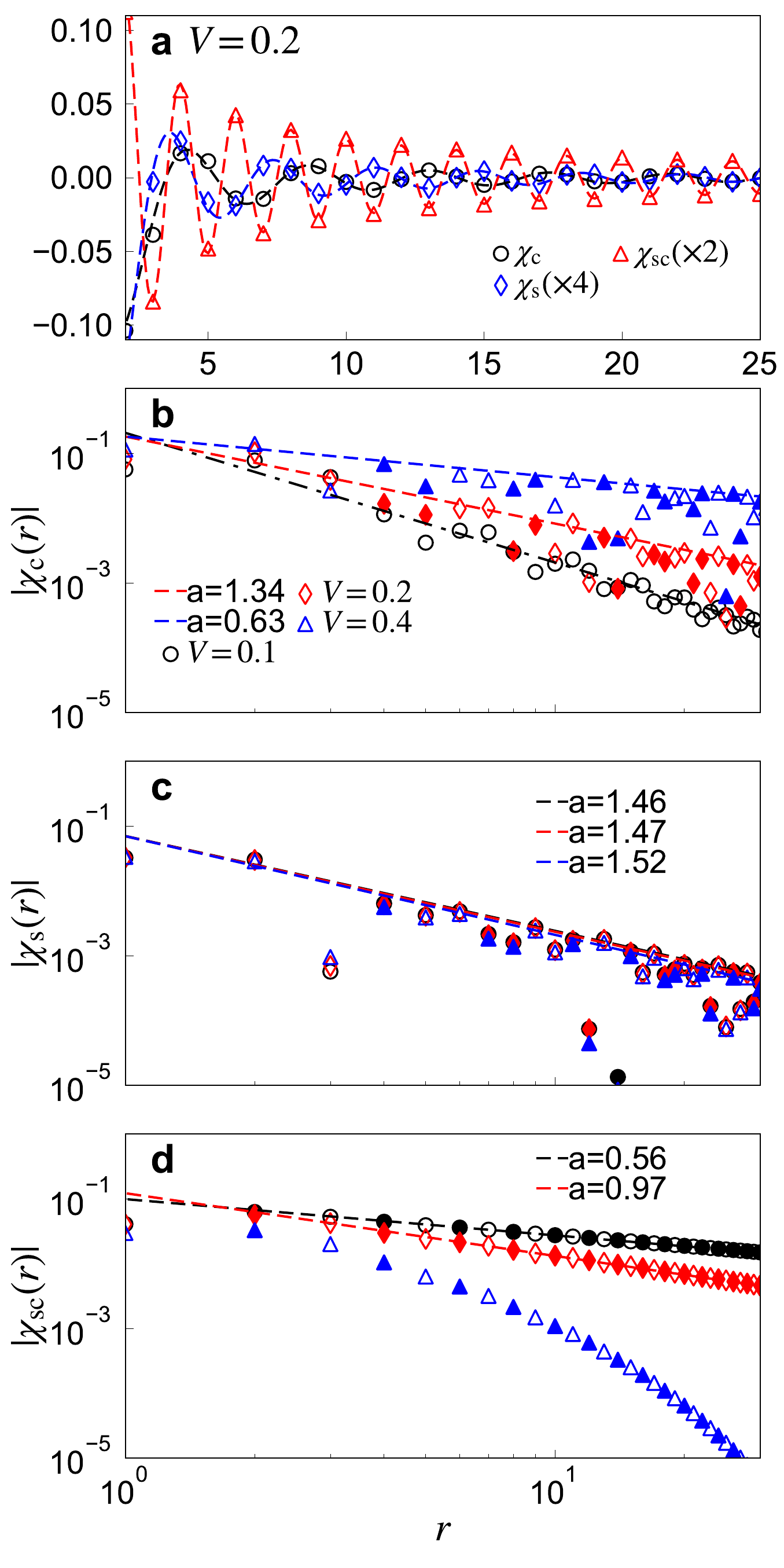} 
  \caption{Correlation functions $\chi$ evaluated by iTEBD for the photo-doped states described by $\hH_{\rm eff2}$.  {\bf a} Normal scale plot for $V=0.2$ and the corresponding fitting with $C_1/r^2 + C_2\cos(q r)/r^a$, where $q=2n_d\pi$, $q=(1-2n_d)\pi$ and $q=\pi$ for charge, spin and SC correlations, respectively.  {\bf c}-{\bf d} Log-scale plots of  the absolute value of the correlation functions for specified values of $V$. Empty (filled) markers correspond to $\chi<0$ ($\chi>0$). 
  The dashed (dot-dashed) lines show $C_2 r^{-a}$ ($C_1 r^{-2}$ ) extracted by fitting.
  For all panels, we use $n_d=0.23$ and $r\in[6,30]$ for the fitting range.
 }
  \label{fig:iTEBD_corr}
\end{figure}

 In Fig.~\ref{fig:iTEBD_phase}, we show the computed nonequilibrium phase diagram for the photo-doped Mott insulator.
 In one-dimensional quantum systems, spatial equal-time correlations can show quasi-long-range order, i.e., power-law decay with a critical exponent $a$ less than 2,
which corresponds to a diverging susceptibility in the low-frequency limit~\cite{Giamarchi_book}.
The corresponding spatial dependence of the correlation functions is shown in Fig.~\ref{fig:iTEBD_corr}. 
We see that generically more than one correlation function exhibits quasi-long-ranged order. 
The phase shown in Fig.~\ref{fig:iTEBD_phase} is identified from the correlation function with the smallest critical exponent.
Without photo-doping, a SDW phase with staggered spin correlations is found \footnote{Because of the SU(2) spin symmetry, there may be a logarithmic correction on top of the power-law decay for the spin correlation. In our results, fits with and without this correction work equally well. The latter fit yields larger $a$, hence the phase diagram and the qualitative argument remains unaffected.}.  
However, other correlations quickly become dominant with photo-doping.
When $V\lesssim 0.2$ $(=\frac{J_{\rm ex}}{2})$, the $\eta$-SC phase emerges in a wide photo-doping range.
This is consistent with recent dynamical mean-field theory (DMFT) analyses for the pure Hubbard model in the infinite spatial dimension employing entropy cooling or heat baths~\cite{Werner2019PRB,Jiajun2020PRB}.
Importantly, the sign of the SC correlations remains staggered regardless of doping and $V$. 
For larger $V$, the CDW phase is stabilized.
We note that, in the extreme photo-doping limit ($n_d = 0.5$), the effective model ($\hH_{\rm dh,ex}+\hH_V$) becomes equivalent to the XXZ model ~\cite{Rosch2008PRL}.
Namely, we have 
\eqq{
  \hH_{\rm eff} & = J_{\rm XY}\sum_{\langle i,j\rangle}  [\heta^x_i \heta^x_{j} + \heta^y_i \heta^y_{j}] + J_{\rm Z} \sum_{\langle i,j\rangle} \heta^z_i \heta^z_{j},
}
where $J_{\rm XY}=-J_{\rm ex}$, $J_{\rm Z}=-J_{\rm ex}+4V$ and $\heta$ represents the $\eta$-operators (see {\bf METHODS}).
The XXZ model shows XY order (quasi-long range order) for  $|J_{\rm XY} |>J_{\rm Z}$, while it shows Ising order (long range order) for $|J_{\rm XY}|<J_{\rm Z}$.
In our language, the former corresponds to $\eta$-SC and the latter to CDW.
Thus, it is natural that the phase transition between  $\eta$-SC and CDW occurs at $|J_{\rm XY}|=J_Z$, i.e. $V=\frac{J_{\rm ex}}{2}$, for strong photo-doping.
Interestingly, the phase boundary remains located near this value over a wide photo-doping range, see Fig.~\ref{fig:iTEBD_phase}.

As Fig.~\ref{fig:iTEBD_corr} shows, more than one order can be quasi-long ranged for a given set of parameters.
When $V$ is small,  the SC correlations are dominant. 
The spin correlations are also quasi-long-ranged, while the charge correlations show no sign of CDW (alternation of signs).
When $V$ is increased, the exponent of the spin correlations remains almost unchanged, while the decay of the SC correlations becomes faster and the CDW correlation starts to develop around $V\gtrsim 0.1$ $(= \frac{J_{\rm ex}}{4})$. 
$V=0.2$ $(=\frac{J_{\rm ex}}{2})$ is in a coexistence regime where  CDW, SDW and $\eta$-SC orders are simultaneously quasi-long ranged [Fig.~\ref{fig:iTEBD_corr}{\bf a}]. While the precise boundaries of the coexistence regime are difficult to determine (see SM), by $V=0.4$ $(=J_{\rm ex})$,  the CDW correlations become dominant and the SC correlations decay exponentially.

\begin{figure}[t]
  \centering
    \hspace{-0.cm}
    \vspace{0.0cm}
\includegraphics[width=87mm]{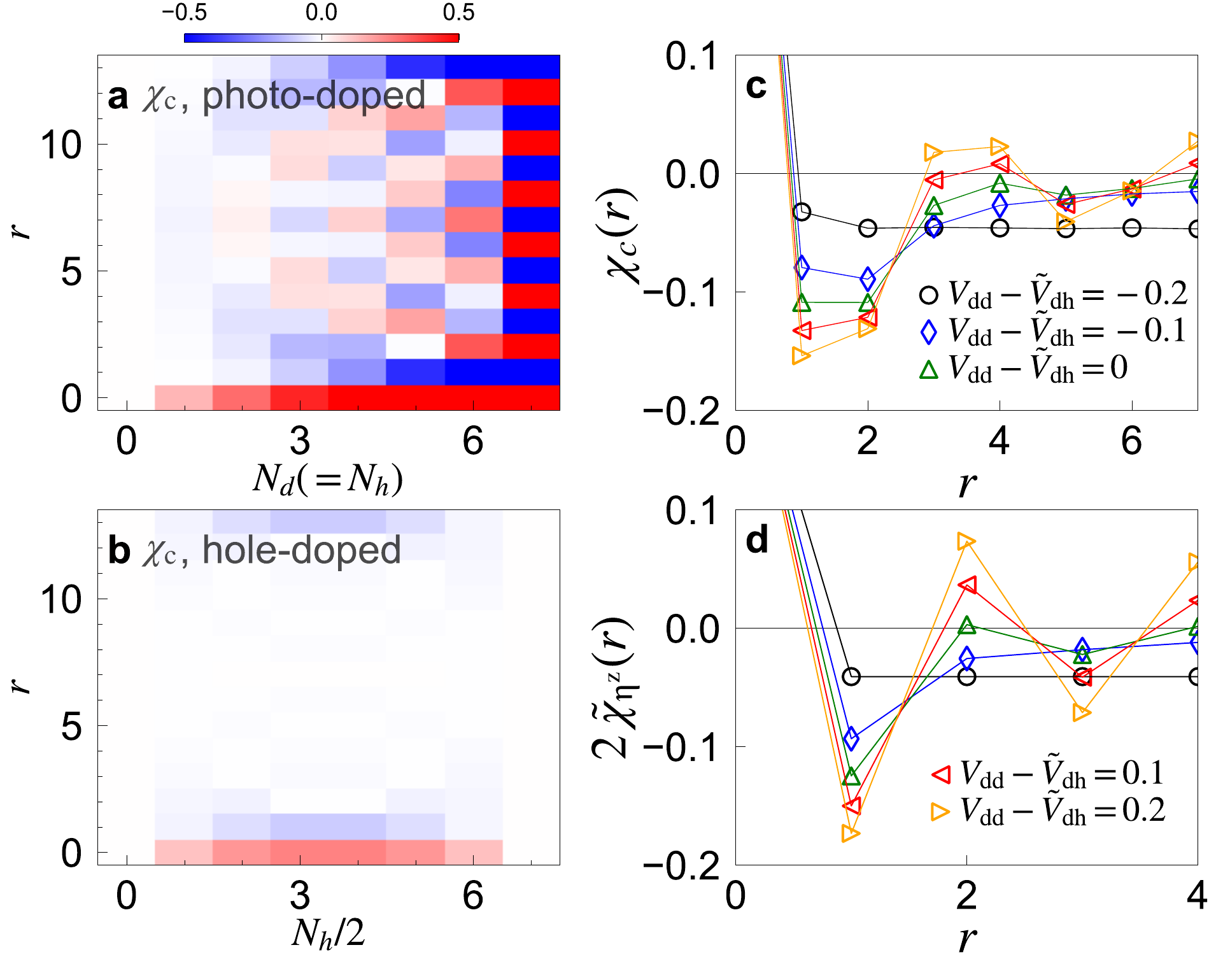} 
  \caption{(a-b) Charge correlation function $\chi_{\rm c}(r)$ for the photo-doped and hole-doped states described by $\hH_{\rm eff2}$. Here, $N_{d(h)}\equiv\langle \hN_{d(h)} \rangle$. 
  {\bf c}-{\bf d}  Dependence of $\chi_{\rm c}(r)$ {\bf c} and $\tilde{\chi}_{\rm \eta^z}(r)$ {\bf d} on the relative magnitude of $V_{\rm dd}$ and $\tilde{V}_{\rm dh}$ 
  for states described by  $\hH_{\rm eff2}+\hat{H}_{V_{\rm dh}}$. 
  Here, we use ED for $N=14$ and  $N_d=N_h=4$  in {\bf c}{\bf d}.
   In all panels, $V=0.4$.
   }
  \label{fig:ED_simple_corr_vs_hole}
\end{figure}

\noindent {\bf Origin and properties of photo-doped phases.} 
The photo-doped states exhibit unique properties.
Firstly, the $\eta$-SC is absent in equilibrium, since in chemically-doped systems either doublons or holons are introduced and hence $\chi_{sc}(r)$ vanishes.
On the other hand, one expects that even in chemically-doped states, CDWs can develop due to the instability of the Fermi surface.
Figures~\ref{fig:ED_simple_corr_vs_hole}{\bf a}{\bf b} however show that the CDW correlations are much stronger in photo-doped than in chemically-doped states.
Furthermore, the CDW correlations show incommensurate oscillations with $q=2n_d\pi$, see Fig.~\ref{fig:iTEBD_corr}{\bf a}.
This indicates that holons and doublons do not bind in pairs. Instead, the holons (doublons) are located in the middle of neighboring doublons (holons), even though the doublon-holon interaction $V_{\rm dh}\equiv \frac{J_{\rm ex}}{4} - V$ is attractive, see $\hH_{\rm dh,ex}+\hH_V$. The absence of binding is also directly confirmed by the evaluation of the doublon-holon correlations (SM). 
This situation is in stark contrast with semiconductors, where the attractive interaction between photo-doped holes and electrons leads to condensation of electron-hole pairs (excitons) at low temperatures~\cite{Keldish1965,Rice1967EI,Perfetto2019PRM,Murakami2020PRB}.

Let us discuss in more detail the physical origin of the CDW phase. 
In the extended Hubbard model the interaction between doublons (and between holons) 
is $V_{\rm dd}\equiv -\frac{J_{\rm ex}}{4} + V$, and thus $V_{\rm dd} \equiv -V_{\rm dh}$. 
To investigate how the interactions among doublons and holons affects the CDW formation, 
we artificially add an interaction between neighboring doublons and holons, $\hat{H}_{V_{\rm dh}}\equiv \Delta V_{\rm dh}\sum_i [\hn_{i,d} \hn_{i+1,h} + \hn_{i,h} \hn_{i+1,d}]$, so that  the doublon-holon interaction becomes  $\tilde{V}_{\rm dh}\equiv V_{\rm dh} + \Delta V_{\rm dh}$. We choose the parameters such that both $\tilde{V}_{\rm dh}$ and $V_{\rm dd}$ are repulsive.  
Figure~\ref{fig:ED_simple_corr_vs_hole}{\bf c} shows that the relative magnitude of the doublon-doublon (holon-holon) and doublon-holon interaction controls the physics and that attractive doublon-holon interaction is not essential for the CDW.
Namely, oscillations in $\chi_c$ appear if $V_{\rm dd}\gtrsim \tilde{V}_{\rm dh}$. 
This indicates that CDW correlations develop between the doublons and holons as if no singlons existed between them. 
The situation is analogous to the spin correlations in the one-dimensional $t$-$J$ model, which can be explained by the squeezed Heisenberg chain without doublons and holons~\cite{Ogata1990PRB,Feng1994PRB}.
Underlying this phenomenon in the one-dimensional $t$-$J$ model is the conservation of the spin configuration in the $J\rightarrow 0$ limit~\cite{Ogata1990PRB}.
Since a singlon always encounters the same neighbors, the system favors the spin configurations described by the Heisenberg hamiltonian.
The same situation is realized in the photo-doped case.  In the limit of $J_{\rm ex}\rightarrow 0$, the configuration of doublons and holons is also conserved due to their peculiar kinematics, see $\hH_{\rm kin,holon} +  \hH_{\rm kin,doub}$. 
(Note that in normal semiconductors, holes and electrons can switch position even in the one-dimensional case.)
Thus, the configurations of doublons and holons are determined by the interaction term $\hH_{\rm dh,ex}+\hH_V$, as in the case of spin configurations in the  $t$-$J$ model.  

To confirm the above scenario, we evaluate the correlations between the doublons and holons in terms of a reduced distance which ignores singlons. The corresponding correlation function is defined as $\tilde{\chi}_{\eta_z}(r) = \langle \sum_{l\geq 0} Q_l \heta^z_{r+l} \heta^z_0 Q_l \rangle$~\cite{Takahashi2005PRB}, where $Q_l$ 
is the projection to states with $l$ singlons between the 0th site and the $(r+l)$th site and $\heta^z_i = \frac{1}{2} (\hn_{i,d}-\hn_{i,h})$ [Fig.~\ref{fig:ED_simple_corr_vs_hole}{\bf d}].
Staggered correlations appear for $V_{\rm dd} \gtrsim \tilde{V}_{\rm dh}$, which supports the above argument.
Note that, even when doublons and holons show the Ising-type order in the squeezed space, the correlations can still exhibit a power-low \cite{Shiba1991PRB}.
We thus conclude that the photo-induced CDW originates from the 
less repulsive doublon-holon interaction  (compared to interactions between the same species), the peculiar kinematics of carriers and the 
one-dimensional configuration. 
Furthermore, the development of correlations between the doublons and holons in the squeezed system without singlons should also apply to systems with $V\lesssim \frac{J_{\rm ex}}{2}$.
In these cases,  the $X$ and $Y$ components of $\hH_{\rm dh,ex}+\hH_V$ (we regard $\hH_{\rm dh,ex}+\hH_V$ as an XXZ model, as in the extreme photo-doing limit) is dominant  and the $\eta$-paring phase emerges. 
This naturally explains the observation that the boundary between the $\eta$-paring phase and the CDW phase is close to  $V\simeq \frac{J_{\rm ex}}{2}$ independent of the photo-doping level.

 \begin{figure}[t]
  \centering
    \hspace{-0.cm}
    \vspace{0.0cm}
\includegraphics[width=87mm]{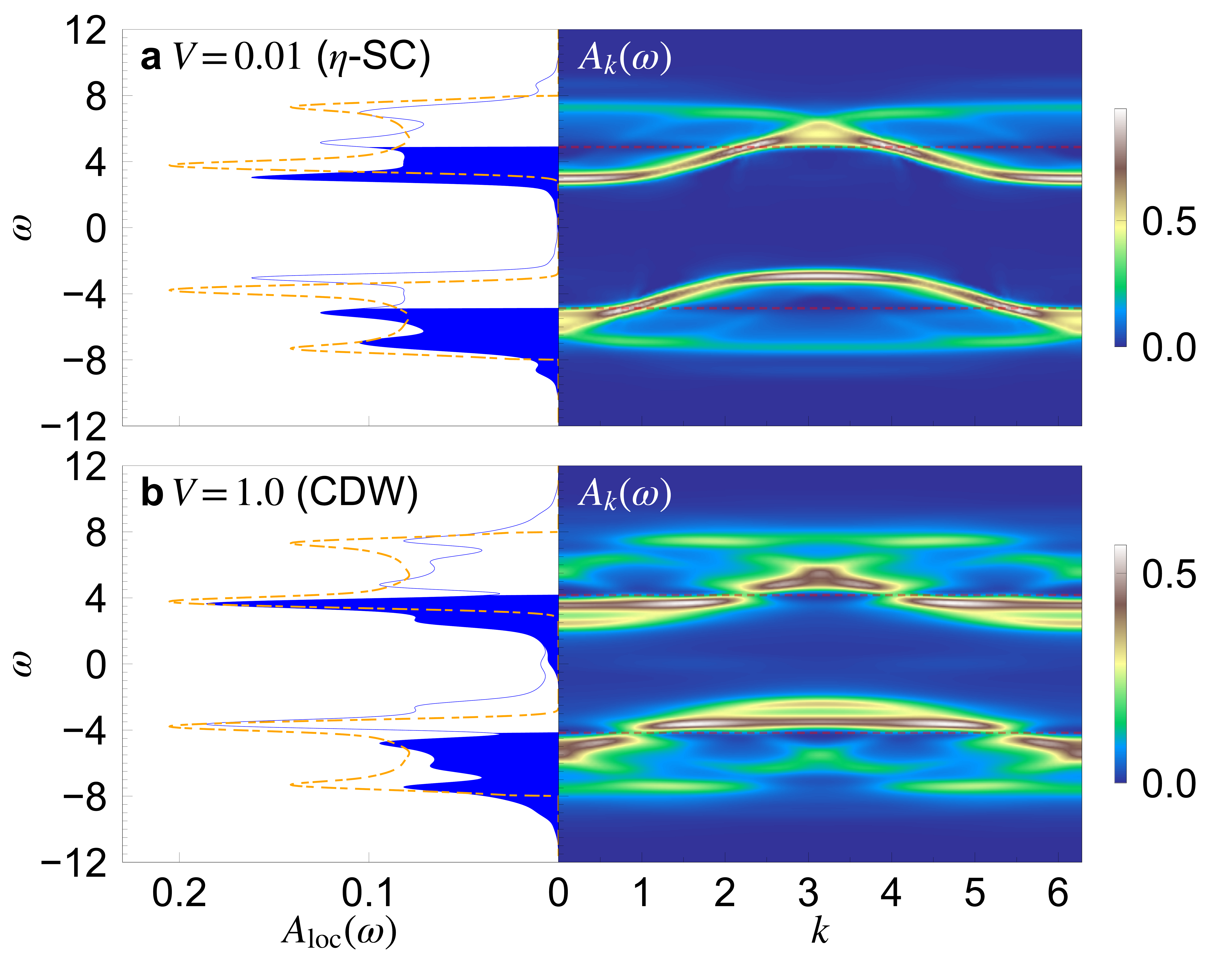} 
  \caption{iTEBD single-particle spectral functions, $A_{\rm loc}(\omega)$ (left) and $A_{k}(\omega)$ (right), for the photo-doped states described by $\hH_{\rm eff2}$. 
  In the left panels, the filled regions indicate the occupied states and dot-dashed lines show $A_{\rm loc}(\omega)$ for the equilibrium (non photo-doped) states.
  In the right panels, dashed lines indicate the Fermi levels in the UHB and LHB. Here, $n_d=0.23$. 
  }
  \label{fig:iTEBD_spectrum}
\end{figure}

\noindent {\bf Single-particle spectra.} 
We now focus on the single-particle spectra, to clarify characteristic features of the different phases.
Figure~\ref{fig:iTEBD_spectrum} shows the momentum-integrated spectrum $A_{\rm loc}(\omega)$ and  the momentum-resolved spectrum $A_{k}(\omega)$ for the $\eta$-SC phase and the CDW phase~\cite{Murakami2021PRB}. 
For the CDW phase, we use $V=1$ to enhance the characteristic features. 
Unlike in equilibrium, but similar to photo-doped semiconductors, the photo-doped system exhibits two “Fermi levels” separating occupied (electron removal spectrum) from unoccupied (electron addition spectrum) states.
The occupied states in the upper Hubbard band (UHB) region correspond to the removal of a doublon, while those in the lower Hubbard band (LHB) region correspond to adding a holon (see Methods for precise definitions).
In the $\eta$-SC phase, within our numerical accuracy, no gap signature appears in $A_{k}(\omega)$  around the new Fermi levels [Fig.~\ref{fig:iTEBD_spectrum}{\bf a}], which is in stark contrast to a normal superconductor with a gap around the Fermi level. 
The absence of a gap is also found for $\eta$-paring states in higher dimensions~\cite{Jiajun2020proc}, and this suggests that the $\eta$-SC state is a kind of gapless superconductivity. 
On the other hand, in the CDW phase, gaps appear at the new Fermi levels  [Fig.~\ref{fig:iTEBD_spectrum}{\bf b}], as in the excitonic phase in photo-doped semiconductors~\cite{Perfetto2019PRM}.

Finally, we observe that in-gap states between the UHB and LHB develop with photo-doping, which are more prominent for larger $V$ [Fig.~\ref{fig:iTEBD_spectrum}].
These states may enable recombination processes  suggesting that our assumption of approximately conserved doublon and holon numbers may become less valid as the excitation density increases.
However, one needs to keep in mind the following points: 
i) For large enough $U$, the Mott gap remains clear and the doublon and holon numbers are approximately conserved.
Since the CDW is driven by $V$, the value of $U$ does not affect its existence and spectral features.
ii) Even when in-gap states develop, the effective equilibrium description is meaningful.  
The recombination rate for a given state can be estimated by Fermi's golden rule, and if this rate is small compared to the intraband relaxation, the transient state can be described by (time-dependent) effective temperatures and chemical potentials~\cite{Florian2018PRB}.
Hence, our results show that the effective equilibrium description can be useful to study the closure/shrinking of a Mott gap via photo-doping, as a result of screened interactions and photo-induced spectral features~\cite{Denis2015PRB}.

\noindent{\bf Discussion}\\
We introduced a GGE-type effective equilibrium description for photo-doped strongly correlated systems. This provides a theoretical framework for systematic studies of nonthermal phases.
Using this effective equilibrium description, we revealed emerging phases in the photo-doped one-dimensional extended Hubbard model. 
The $\eta$-pairing phase is stabilized in the small $V$ regime even when the SU$_c(2)$ symmetry that protects $\eta$-pairing in the pure Hubbard model is absent, and it is characterized by gapless spectra.
The CDW phase emerges in the larger $V$ regime, and it is characterized by gapped spectra.
These states are unique to photo-doped strongly correlated system, where the peculiar kinematics of doublons and holons stabilizes them in a wide doping range.
The similarity between the GGE-type description for strongly correlated systems and the pseudoequilibrium description for the photo-doped semiconductors allowed us to clarify some fundamental differences between these two systems.
In particular, our results demonstrate that photo-doped strongly correlated systems and semiconductors exhibit qualitatively different phases due to the different nature of the injected carriers.
Target systems to look for the characteristic Mott features include candidate materials of one-dimensional Mott insulators ranging from organic crystals, e.g., ET-F$_2$TCNQ, to cuprates, e.g. Sr$_2$CuO$_3$, as well as cold-atom systems.

We also note that further insights into our results for the one-dimensional system may be obtained by using equilibrium concepts such as Luttinger liquid theory and the exact wave function for $J_{\rm ex}\rightarrow 0$~\cite{Ogata1990PRB}. 
In the nonequilibrium state, we expect at most three degrees of freedoms: spin, pseudo-spin (consisting of doublon and holon) and charge (position of singlons).
Thus, unlike in the equilibrium case, the maximum value of the conformal charge $c$ would be $3$, which  may be realized in the $\eta$-pairing phase.  A systematic analysis in this direction is under consideration.

The GGE-type description of photo-doped Mott states can be applied to various models and implemented with different equilibrium techniques such as slave boson approaches~\cite{Kotliar1986PRL,Ogata_2008} and variational methods~\cite{Vidal2003PRL,Corboz2010PRB,Misawa2019}.
The description can provide useful insights into experimental findings, e.g. photo-induced SC-like state, as well as theoretical results. 
For example, a recently found metastable orbital order in a photo-doped multi-orbital system can be reasonably explained by an effective equilibrium picture~\cite{Werner2021PRBL}.
Systematic explorations of nonequilibrium phases in higher dimensions and at nonzero effective temperatures with the GGE-type description should be undertaken. 
For example, the GGE-type description allows us to investigate important aspects of nonequilibrium states, such as the screening of the interactions~\cite{Denis2015PRB} or the stability of photo-induced phases against light irradiation~\cite{Tsuji2021arxiv}. 
These questions are interesting topics for future investigations.

\noindent{\bf Methods}\\
\noindent{\bf Infinite time-evolving block decimation.}
The infinite time-evolving block decimation (iTEBD) method expresses the wave function of the system as a matrix product state (MPS), assuming translational invariance~\cite{Vidal2003PRL}.  
iTEBD directly treats the thermodynamic limit and we use cut-off dimensions $D=1000 \sim 3000$  for the MPS to get converged results.
We use the conservation laws for the numbers of spin-up and spin-down electrons at half-filling to improve the efficiency of the calculations. 
Away from half-filling, the trick with the conservation laws cannot be used, which makes the simulations less efficient. 
Thus, we use the exact diagonalization method in Fig.~\ref{fig:ED_simple_corr_vs_hole}.

\noindent{\bf Single-particle spectrum.}
The single-particle spectrum is defined as follows. The local spectrum is  $A_{\rm loc}(\omega) \equiv  -\frac{1}{\pi}{\rm Im} G^R_i(\omega)$ and the momentum-resolved spectrum is $A_{k}(\omega) \equiv  -\frac{1}{\pi}{\rm Im} G^R_k(\omega)$. Here, $G^R_i(\omega)$ ($G^R_k(\omega)$) is the Fourier transform of the retarded Green's function $G^R_i(t)=-i\langle [\hc_{i\sigma}(t),\hc^\dagger_{i\sigma}(0)]_+\rangle$ ($G^R_k(t)=-i\langle [\hc_{k\sigma}(t),\hc^\dagger_{k\sigma}(0)]_+\rangle$).
Note that $\hc_{i\sigma}(t)$ is the Heisenberg representation of $\hc$ in terms of $\hH_{\rm eff}$.
The occupied spectra correspond to $A^<_{\rm loc}(\omega) \equiv  \frac{1}{2\pi}{\rm Im} G^<_i(\omega)$ and $A^<_{k}(\omega) \equiv  \frac{1}{2\pi}{\rm Im} G^<_k(\omega)$, where $G^<$ denotes the lesser part of the Green's functions.
To evaluate these quantities using the effective equilibrium description and iTEBD, we employ the method proposed by some of the authors~\cite{Murakami2021PRB}. 
Namely, we evaluate $G^R(t)$ using an auxiliary band and perform the Fourier transformation with a Gaussian window, $F_{\rm Gauss}(t) = \exp(-\frac{t^2}{2\sigma^2})$, and we use $\sigma=5.0$.
Thus, the broadening of the resultant spectrum is inevitable and a gap much smaller than the broadening cannot be captured. Still,  we checked that no gap signature appears in the $\eta$ paring state for an increased value of $J_{\rm ex}$, where we would expect an increase of the gap (if any).

\noindent{\bf $H_{\rm eff}$ for the $U$-$V$ Hubbard model.} The explicit expressions for the terms in Eq.~\eqref{eq:Heff} are as follows.
The $\mathcal{O}(v)$ terms are given by
\eqq{
\hH_{\rm kin,holon}(t) = -t_{\rm hop}\sum_{\langle i,j\rangle,\sigma} \bar{n}_{i,\bar{\sigma}} (c^\dagger_{i,\sigma}c_{j,\sigma} + h.c.)  \bar{n}_{j,\bar{\sigma}}, \\
\hH_{\rm kin,doublon}(t) = -t_{\rm hop}\sum_{\langle i,j\rangle,\sigma} n_{i,\bar{\sigma}} (c^\dagger_{i,\sigma}c_{j,\sigma} +h.c.)  n_{j,\bar{\sigma}},
}
where $\bar n=1-n$ and $\bar\sigma$ is the opposite spin to $\sigma$.

For the $\mathcal{O}(\frac{t_{\rm hop}^2}{U})$ terms, we introduce the exchange coupling $J_{\rm ex}= \frac{4t_{\rm hop}^2}{U}$.
With this, the spin exchange term becomes 
  \eqq{
\hH_{\rm spin,ex}= J_{\rm ex} \sum_{\langle i,j\rangle} \hat{\bf  s}_i\cdot \hat{\bf  s}_{j} \label{eq:H_spin},
}
where $\hat{{\bf s}} = \hc^\dagger_\alpha \boldsymbol{\sigma}_{\alpha\beta} \hc^\dagger_\beta$ with $\boldsymbol{\sigma}$ denoting the Pauli matrices.
The exchange term for a doublon and a holon on neighboring sites is 
  \eqq{
  \hH_{\rm dh,ex} & = -J_{\rm ex}\sum_{\langle i,j\rangle} [ \heta^x_i \heta^x_{j} + \heta^y_i \heta^y_{j} + \heta^z_i \heta^z_{j}].
  }
   The shift of the local interaction is described by
 \eqq{
  \hH^{(2)}_{\rm U,shift} & =J_{\rm ex} \sum_i (\hn_{i\uparrow}-\tfrac{1}{2})(\hn_{i\downarrow}-\tfrac{1}{2}).
  }
  Here, the superscript "${(2)}$" indicates that the term is order of $\mathcal{O}(\frac{t_{\rm hop}^2}{U})$.
  
 The 3-site term can be expressed as $\hH_{\rm 3-site}\equiv \hH^{(2)}_{\rm kin,holon} +  \hH^{(2)}_{\rm kin,doub} +  \hH^{(2)}_{\rm dh,slide}$.
Here, $ \hH^{(2)}_{\rm kin,holon}$ and $\hH^{(2)}_{\rm kin,doub}$ are correlated hoppings of holons and doublons, while 
$\hH^{(2)}_{\rm dh,slide}$ shifts the position of a doublon and a holon.
Their expressions are 
    \eqq{
  \hH^{(2)}_{\rm kin,holon} 
  &  =\frac{J_{\rm ex}}{4}\sum_{\langle k,i,j \rangle,\sigma} \Bigl[ 
    n_{i,\bar{\sigma}} c_{j,\sigma}\bar{n}_{j,\bar{\sigma}} \bar{n}_{k,\bar{\sigma}} c^\dagger_{k,\sigma} + 
    h.c.
    \Bigl] \\ \nonumber
    & - \frac{J_{\rm ex}}{4} \sum_{\langle k,i,j\rangle,\sigma}  \Bigl[\bar{n}_{k,\sigma}c^\dagger_{k,\bar{\sigma}} c_{i,\bar{\sigma}} c^\dagger_{i,\sigma} c_{j,\sigma} \bar{n}_{j,\bar{\sigma}}
    + h.c. \Bigl],
  }
 \eqq{
  \hH^{(2)}_{\rm kin,doublon}  
&= \frac{J_{\rm ex}}{4}\sum_{\langle k,i,j\rangle,\sigma}  \Bigl[ \bar{n}_{i,\bar{\sigma}} c^\dagger_{j,\sigma} n_{j,\bar{\sigma}} n_{k,\bar{\sigma}} c_{k,\sigma} 
+   h.c.]\\
& - \frac{J_{\rm ex}}{4} \sum_{\langle k,i,j\rangle,\sigma}  \Bigl[ c^\dagger_{i,\bar{\sigma}} n_{k,\sigma} c_{k,\bar{\sigma}} n_{j,\bar{\sigma}} c^\dagger_{j,\sigma} c_{i,\sigma}
  +  h.c.\Bigl],\nonumber
 }
and 
 \eqq{
\hH^{(2)}_{\rm dh,slide}  
  &= \frac{J_{\rm ex}}{4} \sum_{\langle k,i,j\rangle,\sigma} \Bigl[ c_{i\sigma}^\dagger c_{j,\sigma}\bar{n}_{j,\bar{\sigma}} c^\dagger_{i,\bar{\sigma}}  c_{k,\bar{\sigma}}  n_{k,\sigma}
   +   h.c. \Bigl] \\
 &  + \frac{J_{\rm ex}}{4} \sum_{\langle k,i,j\rangle,\sigma}  \Bigl[n_{j,\bar{\sigma}} c_{j\sigma}^\dagger c_{i,\sigma}  \bar{n}_{k,\sigma} c^\dagger_{k,\bar{\sigma}}  c_{i,\bar{\sigma}} 
  + h.c. \Bigl] \nonumber.
 }
Here, $\langle k,i,j \rangle$ means that both of  ($k$, $i$)  and  ($i$, $j$)  are pairs of neighboring sites. 
The sum is over all possible such combinations (without double counting), where we regard $(k,i,j)=(j,i,k)$.

In the evaluation of the physical quantities, we use the operators of the effective model. To be strict, if physical quantities for the original Hamiltonian are to be computed, one also needs to 
take account of corrections from the SW transformation to the operators. However, these corrections are not necessary to see the leading behavior. 
The same strategy is often used in the evaluation of physical quantities for the Heisenberg model or the $t$-$J$ model.
   
\noindent{\bf Data availability}\\
The data that support the findings of this study are available from the corresponding author upon reasonable request.

\noindent{\bf Code availability}\\
The source code for the calculations performed in this work is available from the corresponding authors upon reasonable request.

\normalem
\bibliographystyle{naturemag}
\bibliography{Ref}

\noindent{\bf Acknowledgements}\\
We thank T.~Oka and F.~Sekiguchi for helpful discussions. 
The calculations have been performed on the Beo05 cluster at the University of Fribourg. 
This work is supported by Grant-in-Aid for Scientific Research from JSPS, KAKENHI Grant Nos. JP19K23425 (Y. M.), JP20K14412 (Y. M.), JP20H05265 (Y. M.), JP21H05017 (Y.M.)a, JP21K03412 (S. T.),  JST CREST Grant No. JPMJCR1901 (Y. M.) JPMJCR19T3 (Y. M. and S. T.), Slovenian Research Agency (ARRS) Grant. No. J1-2455 and P1-0044 (D. G.),  and ERC Consolidator Grant No.~724103 (P. W.). 
A.J.M. is  supported in  part by Programmable Quantum Materials, an Energy Frontier Research Center funded by the U.S. Department of Energy (DOE), Office of Science, Basic Energy Sciences (BES), under award DE-SC0019443. The Flatiron Institute is a division of the Simons Foundation.
T.K. was supported by the JSPS Overseas Research Fellowship.\\

\noindent{\bf Authors contributions}\\
Y. M., D. G. and P. W. conceived the project.  P. W. and A. M. supervised the project. Y. M. made all the numerical simulations. Y. M. and S.T. contributed to the iTEBD code. 
T. K., Z. S. and the rest of authors contributes to the discussion the discussion and the writing of the manuscript.\\

\noindent{\bf Competing interests}\\
The authors declare no competing interests.\\


\appendix

\noindent{\bf Supplementary information}\\
\section{General procedure}
In this section, we explain a general procedure to derive effective models and set the chemical potentials for photo-doped strongly correlated systems (SCSs).
As a generic situation, we consider a Hubbard-type Hamiltonian consisting of a local part and a nonlocal part, $\hH = \hH_{\rm loc} + \hH_{\rm nonloc}$,
where $\hH_{\rm loc} = \sum_i \hH_{{\rm loc},i}$ and $\hH_{\rm nonloc}$ includes nonlocal processes such as hoppings and nonlocal interactions.
We represent the eigenstates of the local part  $\hH_{{\rm loc},i}$ at site $i$ by $|\alpha\rangle_i$, with $\alpha = 1,2,\cdots ,N_{\rm loc}$. For example,  we have $N_{\rm loc}=4$ in the single band Hubbard model.
Next, we introduce pseudo-particle operators $\hd^\dagger_{\alpha,i}$ for the local many-body states $|\alpha\rangle_i$, which are useful to formulate the generic and systematic procedure based on the Schrieffer-Wolff (SW) transformation.
Here, $\hd^\dagger_{\alpha,i}$ is a fermionic (bosonic) creation operator if $|\alpha\rangle$ includes an odd (even) number of fermions, and the physical space is defined by $\sum_\alpha \hd^\dagger_{\alpha,i} \hd_{\alpha,i} =1$.
With the help of these operators, one can express the local Hamiltonian as $\hH_{{\rm loc},i} = \sum_{\alpha=1}^{N_{\rm loc}} E_\alpha \hd^\dagger_{\alpha,i} \hd_{\alpha,i}$.

Let us assume that $E_\alpha-E_{\alpha'}$ can take large values of $\mathcal{O}(U)$, while the remaining terms are $\mathcal{O}(t_{\rm hop})$, so that we can treat $\hH_{\rm nonloc}$ perturbatively.
To this end, we classify the local states into several groups, such that the energy difference of the states within the same group is less than $\mathcal{O}(U)$, and the states in the same group have the same number of physical particles.
For example, in the one band Hubbard model, we have three groups.
The first one consists of holons, the second one consists of doublons and the third one consists of singly occupied states.
The number operator for each group $G$ is defined as $\hat{N}_G = \sum_i \sum_{\alpha \in G} \hd^\dagger_{\alpha,i} \hd_{\alpha,i}$.

$ \hH_{\rm nonloc}$, written in terms of the pseudo-particle operators, reads 
\eqq{
\hH_{\rm nonloc} &= \sum_{i,j\atop\alpha_1,\alpha_2,\alpha_3,\alpha_4} t^{\alpha_1,\alpha_2,\alpha_3,\alpha_4}_{{\rm hop},i,j} \hd^\dagger_{i,\alpha_1} \hd_{i,\alpha_2} \hd^\dagger_{j,\alpha_3} \hd_{j,\alpha_4} \nonumber \\
& = \hH_{\rm nonloc,0} + \hH_{\rm nonloc,\pm},
}
where $\hH_{\rm nonloc,0} \equiv \sum'_{i,j,\alpha_1,\alpha_2,\alpha_3,\alpha_4} t^{\alpha_1,\alpha_2,\alpha_3,\alpha_4}_{{\rm hop},i,j}$ $\times\hd^\dagger_{i,\alpha_1} \hd_{i,\alpha_2} \hd^\dagger_{j,\alpha_3} \hd_{j,\alpha_4}$ 
and $ \sum'_{i,j,\alpha_1,\alpha_2,\alpha_3,\alpha_4}$ denotes the restriction of the sum 
to those terms $\hd^\dagger_{i,\alpha_1} \hd_{i,\alpha_2} \hd^\dagger_{j,\alpha_3} \hd_{j,\alpha_4}$ which do not change the number of local states in each group $\{ N_G\}$.
One the other hand, $\hH_{\rm nonloc,\pm} \equiv \sum''_{i,j,\alpha_1,\alpha_2,\alpha_3,\alpha_4} t^{\alpha_1,\alpha_2,\alpha_3,\alpha_4}_{{\rm hop},i,j} \hd^\dagger_{i,\alpha_1} \hd_{i,\alpha_2} \hd^\dagger_{j,\alpha_3} \hd_{j,\alpha_4} $,  
where $ \sum''_{i,j,\alpha_1,\alpha_2,\alpha_3,\alpha_4}$ denotes the sum
 of those terms $\hd^\dagger_{i,\alpha_1} \hd_{i,\alpha_2} \hd^\dagger_{j,\alpha_3} \hd_{j,\alpha_4}$ which change  $\{ N_G\}$.
 Note that here we implicitly assume that, if $\{ N_G\}$ changes, the corresponding energy change  $E_{\alpha_1}-E_{\alpha_2} + E_{\alpha_3} - E_{\alpha_4}$ is $\mathcal{O}(U)$.

Now, we apply the SW transformation, $\hH' = e^{i\hS}\hH e^{-i\hS}$ with $\hS=\sum_{l\geq 1} \hS^{(l)}$.
Here $S^{(l)}$ is $\mathcal{O}(v/U)^l$. $\hS^{(l)}$ is chosen such that, at each order, $\hH'$ does not include processes that change $\{ N_G\}$.
The lowest order term of $\hS$ is chosen such that 
\eqq{
\hH_{{\rm nonloc},\pm} = i[ \hH_{\rm loc},\hS^{(1)}].
}
This is satisfied for
\eqq{
\hS^{(1)} & =-i \sum''_{i,j\atop\alpha_1,\alpha_2,\alpha_3,\alpha_4}  \frac{t^{\alpha_1,\alpha_2,\alpha_3,\alpha_4}_{{\rm hop},i,j}}{E_{\alpha_1} - E_{\alpha_2} + E_{\alpha_3} - E_{\alpha_4}} \nonumber\\
&\hspace{24mm}\times  \hd^\dagger_{i,\alpha_1} \hd_{i,\alpha_2} \hd^\dagger_{j,\alpha_3} \hd_{j,\alpha_4}.
} 
In general, for any operator that is expressed as 
\eqq{
\hH_{\pm} = \sum_{i_1,i_2,\cdots}\sum''_{\alpha_1,\alpha_1',\alpha_2\cdots} t^{\alpha_1,\alpha_1',\alpha_2\cdots}_{{\rm hop},i_1,i_2,\cdots} \hd^\dagger_{i_1,\alpha_1}\hd_{i_1,\alpha'_1}\hd^\dagger_{i_2,\alpha_2}\hd_{i_2,\alpha'_2}\cdots,
}
we can define 
\eqq{
\hS' &=-i  \sum_{i_1,i_2,\cdots}\sum''_{\alpha_1,\alpha_1',\alpha_2\cdots} \frac{t^{\alpha_1,\alpha_1',\alpha_2\cdots}_{{\rm hop},i_1,i_2,\cdots}}{E_{\alpha_1}- E_{\alpha'_1} + E_{\alpha_2}- E_{\alpha'_2} + \cdots}  \nonumber\\
&\hspace{22mm}\times\hd^\dagger_{i_1,\alpha_1}\hd_{i_1,\alpha'_1}\hd^\dagger_{i_2,\alpha_2}\hd_{i_2,\alpha'_2}\cdots
} 
and one finds that it satisfies 
\eqq{
\hH_{\pm} = i[ \hH_{\rm loc},\hS'].
}
This fact can be used to obtain the higher order corrections for $\hS$.

To summarize, the effective Hamiltonian up to second order becomes
\eqq{
\hH_{\rm eff} &= \hH_{\rm loc} + \hH_{\rm nonloc,0} + i[ \hH_{\rm nonloc,\pm} , \hS^{(1)}] \Bigl |_{ 0} .
}
Here, $|_{ 0}$ means that we only consider terms that do not change $\{ N_G\}$.
In other words, $\hH_{\rm eff}$ commutes with $\hat{N}_G$ and recombination processes are explicitly removed.

In a photo-doped system, the number of local states in each group can be considered as approximately conserved when the gap between different bands ($\mathcal{O}(U)$) is large enough.
On the other hand, a relaxation within the energetically separated bands occurs due to the scattering between local states and energy dissipation to other degrees of freedom such as phonons.
For extended systems, the resulting steady state can be described by introducing  ``chemical potentials" $\mu_G$ for each group,
\eqq{
\hK_{\rm eff}  = \hH_{\rm eff}  -\sum_G \mu_G \hat{N}_G.
}
The properties of the nonequilibrium steady state may be described by this grand canonical Hamiltonian $\hK_{\rm eff}$ and an effective inverse temperature $\beta_{\rm eff}$, i.e., by the density matrix $\hrho_{\rm eff} = \exp(-\beta_{\rm eff} \hat{K}_{\rm eff})$.
Now the problem becomes essentially equivalent  to a conventional equilibrium problem, so that one can apply suitable equilibrium techniques to study the nonequilibrium state of SCSs.

Strictly speaking, the above construction implicitly assumes that, within the same group, the ratio of the states can be changed. Such transitions may be caused by (i) hopping processes (i.e. effects of the effective model) or (ii) as a result of some assumed coupling with the environment. However, when neither of the two conditions applies, one needs to find a subgroup of $G$  and introduce a chemical potential for each subgroup. (The remaining procedure is the same as above.) One may encounter such a case for example in multi-orbital problems with degenerated local states. The specific strategy will depend on the model and the likely properties of the environment. 

\section{Response functions}
Static observables can be evaluated directly using  $\hrho_{\rm eff} = \exp(-\beta_{\rm eff} \hat{K}_{\rm eff})$.
However, for response functions, one has to be careful since the time evolution involved is described by $\hH_{\rm eff}$ and not by $\hat{K}_{\rm eff}$.
Specifically, to calculate response functions, we need to evaluate quantities of the type
\eqq{
\tilde{F}_{AB}(t) = \frac{1}{Z_{\rm eff}} \text{Tr}\big[\hrho_{\rm eff}  e^{i \hH_{\rm eff}  t} \hat{A} e^{-i\hH_{\rm eff}  t} \hat{B} \big],
}
where $Z_{\rm eff} = \text{Tr}[\hrho_{\rm eff}] $.
To evaluate this, we express $\hA$ and $\hB$ as 
\eqq{
\hA = \sum_\alpha \hA_\alpha, \;\;  \hB = \sum_\alpha \hB_\alpha,
}
where 
$\hA_\alpha$ (and $\hB_\alpha$) changes the number of states in each group ($G$) by $\Delta N_{G,\alpha}$.
The operator hence 
changes the expectation value of $\hH_\mu\equiv  -\sum_G \mu_G \hat{N}_G$  by $\lambda_\alpha \equiv   -\sum_G \mu_G \Delta N_{G,\alpha}$.
Therefore, we have 
\eqq{
e^{-i\hH_\mu t}\hA_\alpha e^{i\hH_\mu t} = e^{-i\lambda_\alpha t}\hA_\alpha.
}
We thus obtain the following expression for $\tilde{F}_{AB}$:
\eqq{
\tilde{F}_{AB}(t)
& = \sum_\alpha  e^{i\lambda_\alpha t} \frac{1}{Z_{\rm eff}} \text{Tr}[\hrho_{\rm eff} e^{i \hat{K}_{\rm eff} t}   \hat{A}_\alpha   e^{-i\hat{K}_{\rm eff} t} \hat{B}_{-\alpha} ].
}
Here, $-\alpha$ means that  $\Delta N_{G,-\alpha} = -\Delta N_{G,\alpha}$.
The last equation implies that one can evaluate $ \frac{1}{Z_{\rm eff}} \text{Tr}[\hrho_{\rm eff} e^{i \hat{K}_{\rm eff} t}   \hat{A}_\alpha   e^{-i\hat{K}_{\rm eff} t} \hat{B}_{-\alpha} ]$ using some equilibrium formalism and the ``Hamiltonian" $\hat{K}_{\rm eff}$. To obtain the response function, one needs to multiply these results by $e^{i\lambda_\alpha t}$ and sum over $\alpha$.

In the main text, we use the above formalism to evaluate the single-particle spectra for the photo-doped systems with iTEBD \cite{Murakami2021PRB}. 
This indeed yields consistent results with ED for finite systems.



 \begin{figure}[t]
  \centering
    \hspace{-0.cm}
    \vspace{0.0cm}
\includegraphics[width=60mm]{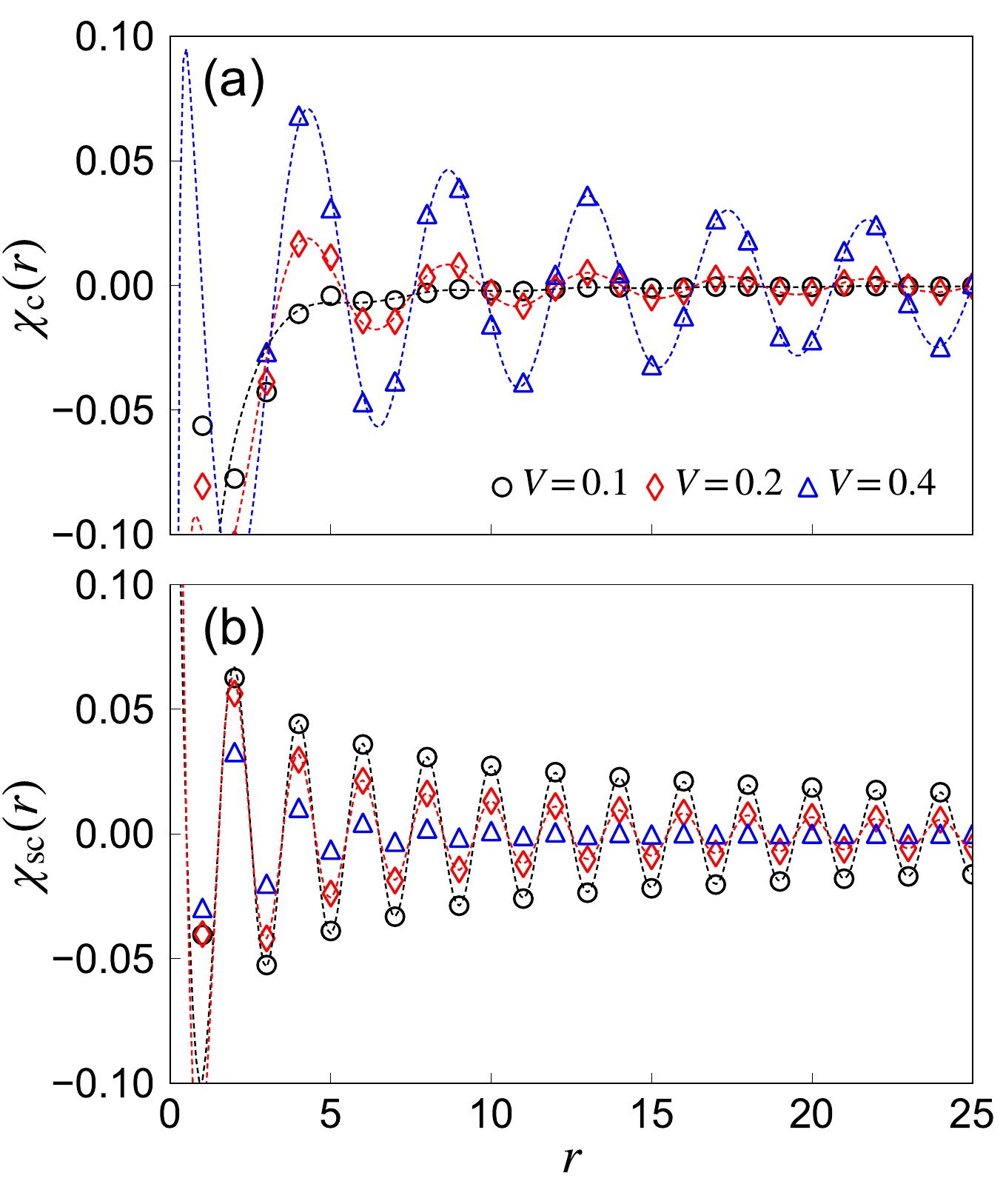} 
  \caption{Correlation functions $\chi$ evaluated by iTEBD for the photo-doped states described by $\hH_{\rm eff2}$.  The data sets are the same as for Fig.~2 in the main text. 
  The dashed lines shows $C_1/r^2 + C_2\cos(q r)/r^a$ extracted by fitting for $r\in[6,30]$, where $q=2n_d\pi$ and $q=\pi$ for charge and SC correlations, respectively. }
  \label{fig:iTEBD_corr2}
\end{figure}

 \begin{figure}[b]
  \centering
    \hspace{-0.cm}
    \vspace{0.0cm}
\includegraphics[width=85mm]{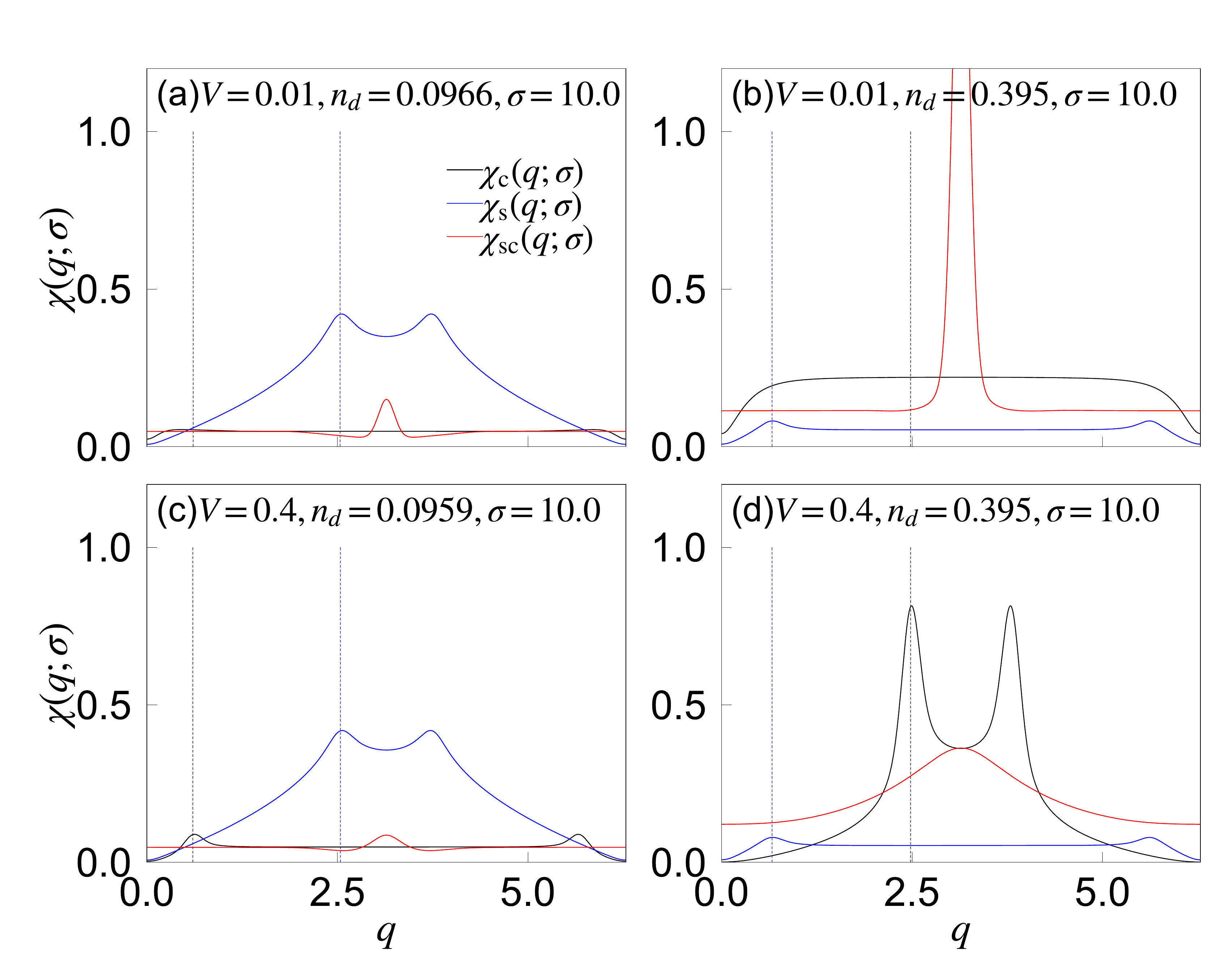} 
  \caption{Window-Fourier transformed correlations $\chi(q;\sigma)$ for the indicated conditions. Here, we use $\hH_{\rm eff2}$, half filling for $U=10$ and $T_{\rm eff}=0$. 
  Vertical dashed lines show $q=(1-2n_d)\pi$ (blue) and $q=2n_d\pi$ (black). 	
  }
  \label{fig:iTEBD_chi_q_windowed}
\end{figure}
 
  \section{Additional results from iTEBD}
 In Fig.~\ref{fig:iTEBD_corr2}, we show the correlation functions as a normal scale plot. 
  The data sets are the same as for Fig.~2 in the main text.
  One can now see how the correlation functions oscillate and how the fitting works.
  In order to further clarify  the size of the correlations in the short range and the relevant momentum,
  we consider the window-Fourier transformation of the correlation functions $\chi(q;\sigma)$
  defined as 
  \eqq{
  \chi_{c}(q;\sigma)  &= \alpha_c \sum_r e^{iqr}  \chi_c(r) F_{\rm gauss}(r;\sigma),\\
    \chi_{s}(q;\sigma)  &= \alpha_s \sum_r e^{iqr}  \chi_s(r) F_{\rm gauss}(r;\sigma),\\
    \chi_{sc}(q;\sigma)  &= \alpha_{\rm sc} \sum_r e^{iqr}  \chi_{\rm sc}(r) F_{\rm gauss}(r;\sigma).
  }
  Here, $F_{\rm gauss}(r;\sigma)\equiv \exp(-\frac{r^2}{2\sigma^2})$. 
  We introduced $\alpha_c=\frac{1}{4}$, $\alpha_s=1$ and $\alpha_{\rm sc}=\frac{1}{2}$, so that the direct comparison of the intensities of different correlations is meaningful.
 Note that $\chi_c(r)\sim4\eta^z\eta^z$ and $\chi_{\rm sc}\sim\eta^x\eta^x + \eta^y\eta^y$.
 
 In Fig.~\ref{fig:iTEBD_chi_q_windowed}, we show $\chi(q;\sigma)$ for several cases.
 $\chi_{sc}(q;\sigma)$ always exhibits a peak at $q=\pi$, i.e. staggered SC correlations.
 $\chi_{s}(q;\sigma)$ also shows peaks, which are located at  $q=(1-2n_d)\pi$ and  $q=(1+2n_d)\pi$.
 When $V$ is small,  $\chi_{c}(q;\sigma)$ shows no clear peak, which is consistent with the lack of CDW correlations in $\chi_c(r)$.
 When $V$ is sufficiently large, peaks develop at $q=2n_d\pi$ and  $q=(2-2n_d)\pi$.
  
\begin{figure}[t]
  \centering
    \hspace{-0.cm}
    \vspace{0.0cm}
\includegraphics[width=65mm]{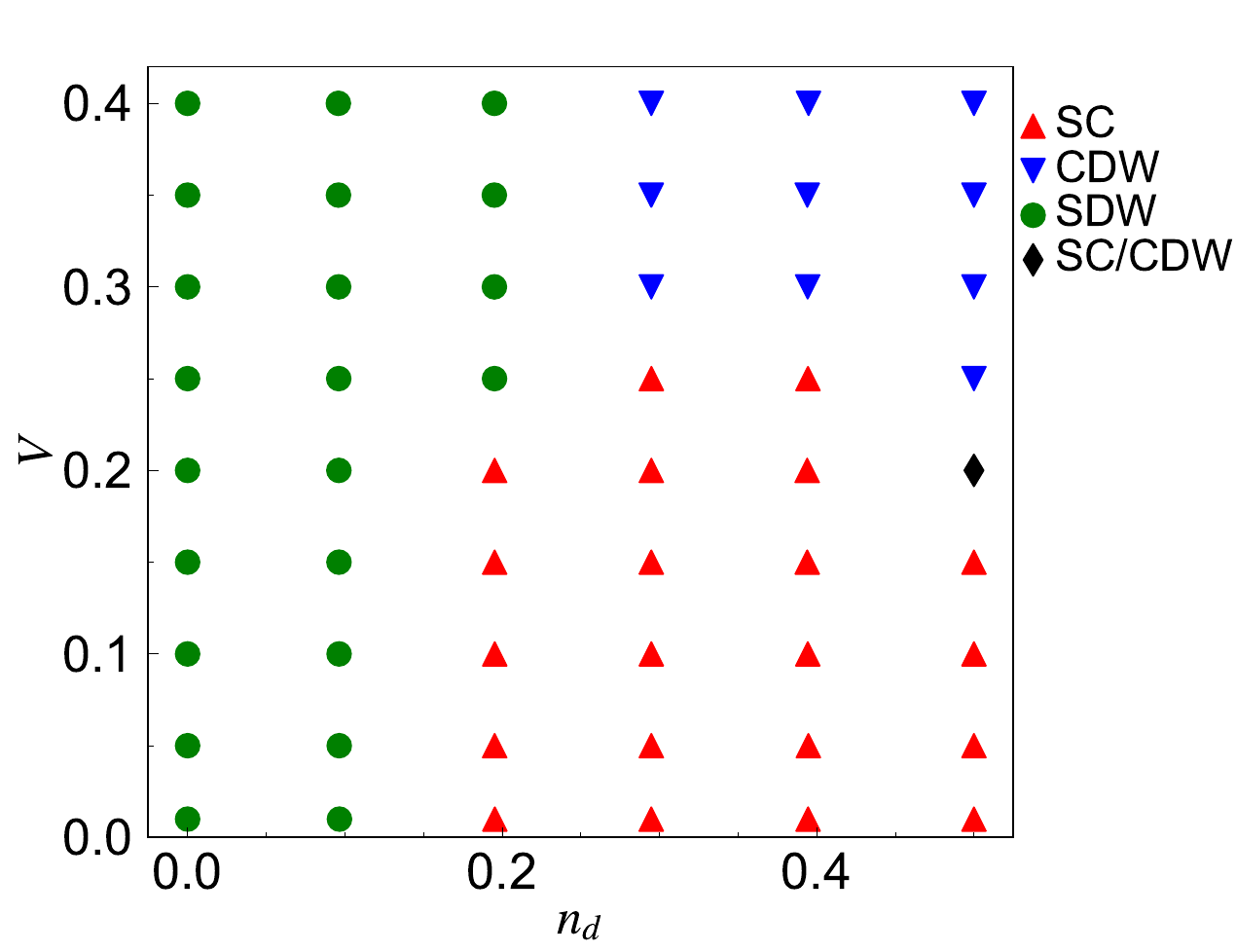} 
  \caption{iTEBD Phase diagram of the photo-doped Mott insulator determined by the relative size of the window-Fourier transformed correlations $\chi(q;\sigma)$ at the peak position. Here, we use $\hH_{\rm eff2}$, 
  half filling, $U=10$ and $T_{\rm eff}=0$. $\sigma = 10.0$ is used for the window.}
  \label{fig:iTEBD_phase2}
\end{figure}

 The window-Fourier transformed correlations also help us to identify the dominant ``short-range" correlation by comparing the peak intensities of $\chi(q;\sigma)$, see illustration in Fig.~\ref{fig:iTEBD_phase2}. 
 Compared to the phase diagram determined by the exponents of the correlation functions in the main text, the SDW region is extended and the CDW region is suppressed.
 We note that the relative size of peaks strongly depends on the choice of the width of the window $\sigma$.

\begin{figure}[t]
  \centering
    \hspace{-0.cm}
    \vspace{0.0cm}
\includegraphics[width=55mm]{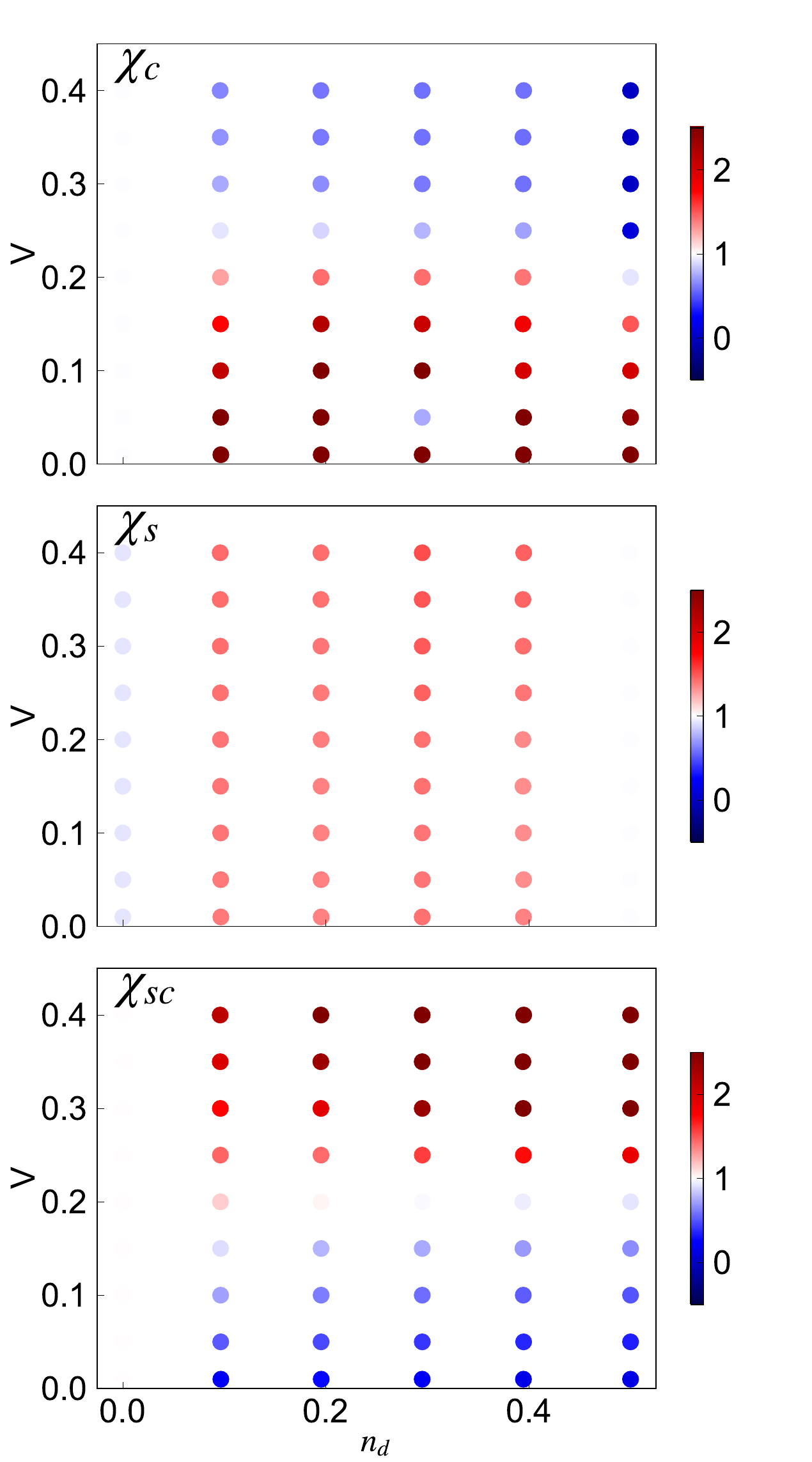} 
  \caption{Critical exponents  extracted by fitting correlation functions ($\chi(r)$) with $C_1/r^2 + C_2\cos(q r)/r^a$, where $q=2n_d\pi$, $q=(1-2n_d)\pi$ and $q=\pi$ for charge, spin and SC correlations, respectively.
  Here we use  iTEBD for $\hH_{\rm eff2}$  at half filling with $U=10$.  We use $r\in[6,30]$ for the fitting range.}
  \label{fig:iTEBD_exponents}
\end{figure}

In Fig.~\ref{fig:iTEBD_exponents}, we show the critical exponents obtained by the fitting of  $\chi(r)$ with $C_1/r^2 + C_2\cos(q r)/r^a$, where $q=2n_d\pi$, $q=(1-2n_d)\pi$ and $q=\pi$ for charge, spin and SC correlations, respectively.
It is clear that for $V=J_{\rm ex}/2$, the exponents of CDW, SDW and SC correlations are all less than 2 (quasi-long ranged).
However, it turns out that determining the exact boundaries where the exponents exceed $2$ is difficult within our present numerical scheme, where the correlation functions can be converged up to $r\simeq 30$ against the cut-off dimension.
Even though the coexistence phase is clearly extended, its boundary fluctuates depending on the fitting range. 
We however confirmed that the relative size of the exponents, which is used for Fig.~1 in the main text, is robust against the fitting range.

\begin{figure}[t]
  \centering
    \hspace{-0.cm}
    \vspace{0.0cm}
\includegraphics[width=55mm]{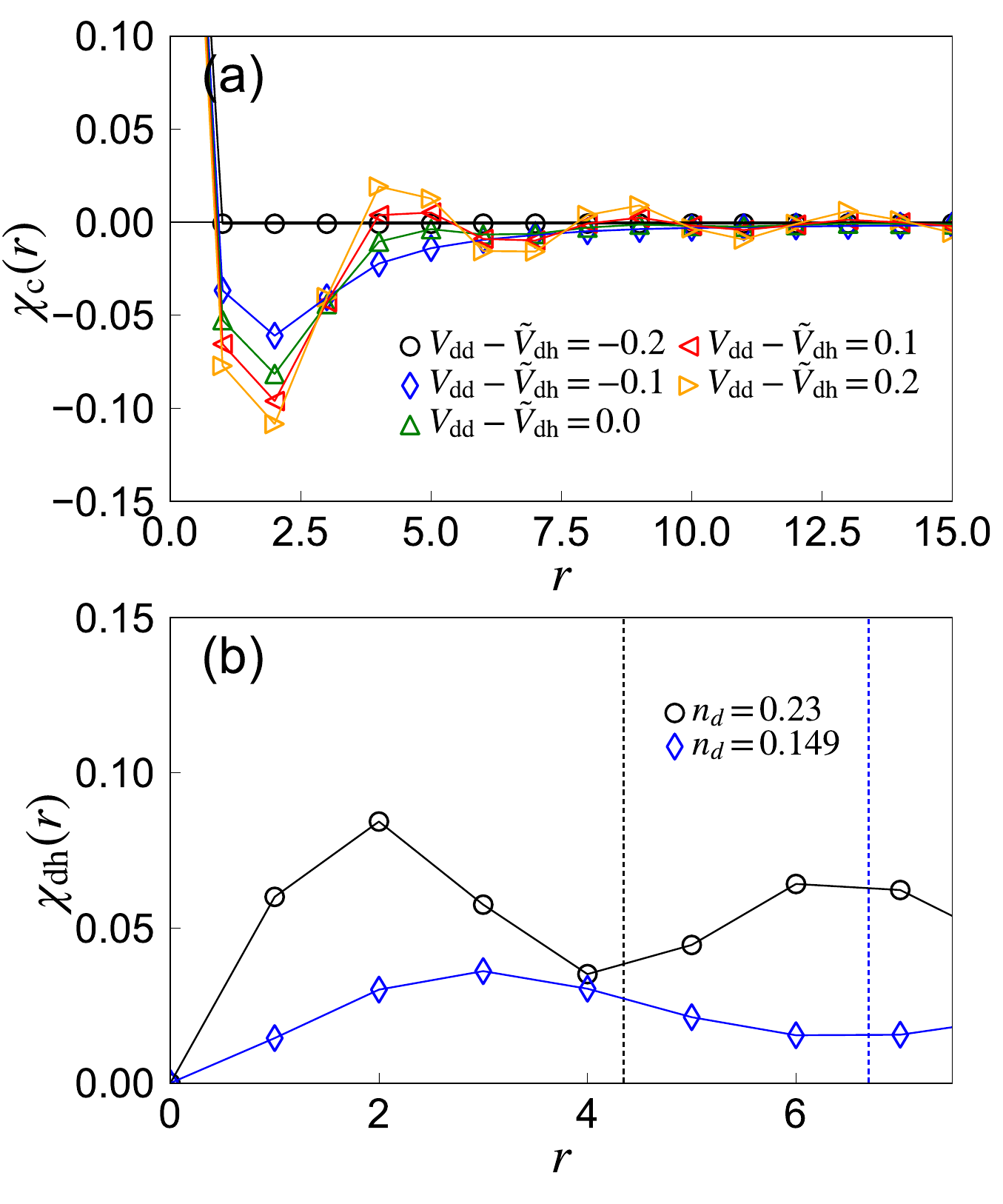} 
  \caption{(a) Dependence of $\chi_{\rm c}(r)$ on the relative magnitude of $V_{\rm dd}$ and $\tilde{V}_{\rm dh}$ for states described by  $\hH_{\rm eff2}+\hat{H}_{V_{\rm dh}}$. Here we use iTEBD and set $n_d=0.23$.
  (b) Doublon-holon correlation function $\chi_{dh}(r)$ for  the CDW phase described by $\hH_{\rm eff2}$. Here, $V=0.4$. The vertical line indicates the mean distance between the doublons, i.e. $1/n_d$.}
  \label{fig:TEBD_simple_comp_holedope_supp}
\end{figure}

In Fig.~\ref{fig:TEBD_simple_comp_holedope_supp}, we show supplemental data for the CDW states. 
Panel (a) shows the results of the dependence on $\hat{H}_{V_{\rm dh}}$, analogous to Fig.~3(c) in the main text which was obtained by ED. 
 As in the ED case, the CDW correlations start to develop when  $\tilde{V}_{\rm dh}<V_{\rm dd}$.
Panel (b) shows the correlation function between the doublons and holons, $\chi_{dh}(r) = \frac{1}{N} \sum_i \langle  \hn_{i+r,h} \hn_{i,d}\rangle$, in the CDW phase.
When a doublon and a holon form a bound pair, the amplitude should be largest at $r=1$. However, the result shows that the location of the holon is just in the middle of neighboring doublons,
which indicates that the doublons and holons are unbound.

\section{Some remarks related to $\eta$ operators} \label{sec:lemmas}
In this section, we make some remarks related to the $\eta$ operators and the effective Hamiltonian for the Hubbard model ($V=0$). 
The following discussion implies that the homogeneous state with the $\eta$-type long range SC correlations is on the verge of phase separation.

At $V=0$, $\hH$ has SU$_c$(2) symmetry. 
Namely,  the following commutation relation with the $\eta$-operators~\cite{Yang1989PRL,Essler2005} holds: $[\hH,\heta^+] = 0, [\hH,\heta^z] = 0$, and $[\hH,\boldsymbol{\heta}^2] = 0$.
Here, $\heta^+ = \sum_j (-)^j \hc_{j\downarrow}^\dagger c_{j\uparrow}^\dagger = \sum_j \eta_j^+,  \;\;  \heta^- = \sum_j (-)^j  \hc_{j\uparrow} \hc_{j\downarrow} = \sum_j \eta_j^-,  \;\; \heta^z = \frac{1}{2}\sum_j (\hn_j -1) = \sum_j \heta^z_j$.
One can show that the SU$_c$(2)  symmetry holds also for $\hH_{\rm eff}$ as well as $\hH_{\rm eff2}$.
We note that these are also true for the grand canonical Hamiltonian $K_{\rm eff}$ at half-filling.

 \subsection{Expectation value of $\eta$ correlations}
 When the Hamiltonian commutes with $\hat{\boldsymbol{\eta}}^2$ and $\heta^z$, 
the three operators have a common basis of eigenstates, which we 
 express as $|\alpha,\eta,\eta^z\rangle_N$. Here, $\hat{\boldsymbol{\eta}}^2|\alpha,\eta,\eta^z\rangle_N = \eta(\eta+1)|\alpha,\eta,\eta^z\rangle_N$, $\heta^z|\alpha,\eta,\eta^z\rangle_N = \eta^z |\alpha,\eta,\eta^z\rangle_N$, $\alpha$ describes the remaining degrees of freedom and $N$ is the system size.
 Then, we have 
 \eqq{
 &\langle \alpha,\eta,\eta^z| \heta^+ \heta^- |\alpha,\eta,\eta^z\rangle_N \nonumber\\
 &= \sum_{j,l}(-)^{j-l} \langle \alpha,\eta,\eta^z| \hc^\dagger_{j,\downarrow}  \hc^\dagger_{j,\uparrow}  \hc_{l,\uparrow}  \hc_{l,\downarrow} |\alpha,\eta,\eta^z\rangle_N   \nonumber\\
 &= (\eta-\eta_z)(\eta+\eta_z+1). \label{eq:eta_eq_3}
 }
When $|\alpha,\eta,\eta^z\rangle_N$ shows long-range $\eta$-type SC correlations, we have $\langle \heta^+ \heta^-\rangle_N=\mathcal{O}( N^2)$.
In such a case, from eq.~\eqref{eq:eta_eq_3}, we have $\eta-\eta_z=\mathcal{O}(N)$ and $ \eta+\eta_z=\mathcal{O}(N)$.

 \subsection{Stability of the $\eta$ states in the effective model}
For the effective model of the Hubbard model ($V=0$) we have  $[\hH_{\rm eff},\heta^\pm]=\pm X\heta^\pm$, and $\hH_{\rm eff}$, $\heta^2$, $\heta^z$, $\hN_d$ and $\hN_h$ commute with each other. 
Here, $X$ is some number and it is zero in the present way of expressing Hamiltonian.
$\heta^z$, $\hN_d$ and $\hN_h$ are not independent, since $\heta^z =\frac{1}{2}(\hN_d-\hN_h)$.
(The following arguments apply to $\hH_{\rm eff2}$ as well.)
Let us assume that a homogeneous state $|\alpha,\eta,N_d,N_h\rangle_N$ is a ground state within a subspace specified by $N_d (=\mathcal{O}(N))$ and $N_h(=\mathcal{O}(N))$. 
We express the energy as $E_0=N\epsilon_0$ and assume $\langle \heta^+ \heta^-\rangle =\mathcal{O}( N^2)$, i.e., an $\eta$-pairing state.
 Now we introduce the following two states,
 \eqq{
 &|\alpha,\eta,N_d+M,N_h-M\rangle_N \propto (\heta^+)^M|\alpha,\eta,N_d,N_h\rangle_N, \nonumber\\
 &  |\alpha,\eta,N_d-M',N_h+M'\rangle_N \propto (\heta^-)^{M'}|\alpha,\eta,N_d,N_h\rangle_N.
 }
 These are also eigenstates of $\hH_{\rm eff}$, whose energies are $E_0+MX$ and $E_0-M'X$, respectively.
Then, we consider the case where a fraction $M'/(M+M')$ of the system is described by  $|\alpha,\eta,N_d+M,N_h-M\rangle_N$ and the rest by $ |\alpha,\eta,N_d-M',N_h+M'\rangle_N$.
 (Strictly speaking, we need to rescale $N_d,N_h,M,N$ by a factor of $M'/(M+M')$ for the former and  $N_d,N_h,M',N$ by a factor of $M/(M+M')$ for the latter.)
Still, the energy per site remains $\epsilon_0$, the density of doublons per site remains $N_d/N$ and for holons per site $N_h/N$.
 Therefore,  $|\alpha,\eta,N_d,N_h\rangle_N$ is thermodynamically unstable in the sense that there exist many inhomogeneous states with the same energy.

 \section{Additional results from ED}
 In this section, we present supplementary results obtained by exact diagonalization (ED) for finite size systems.
 Here we set the system size to $N=14$ and apply the periodic boundary condition.  We use $U=10$ as in the main text.
 For any photo-doping, a homogeneous solution is obtained for finite size systems with ED.
 The ED calculations show that the energy surface is flat along the line $N_d+N_h = {\rm constant}$ for $V=0$ (the pure Hubbard model) as we discussed in the previous section. 
 Furthermore, they confirm the results from iTEBD in the thermodynamic limit.
We note that for ED, one can explicitly specify the doublon number ($N_d$) and  the  holon number ($N_h$), hence there is no need to introduce $\mu_U$.

 \subsection{Stability against phase separation}
 
 \begin{figure}[t]
  \centering
    \hspace{-0.cm}
    \vspace{0.0cm}
\includegraphics[width=70mm]{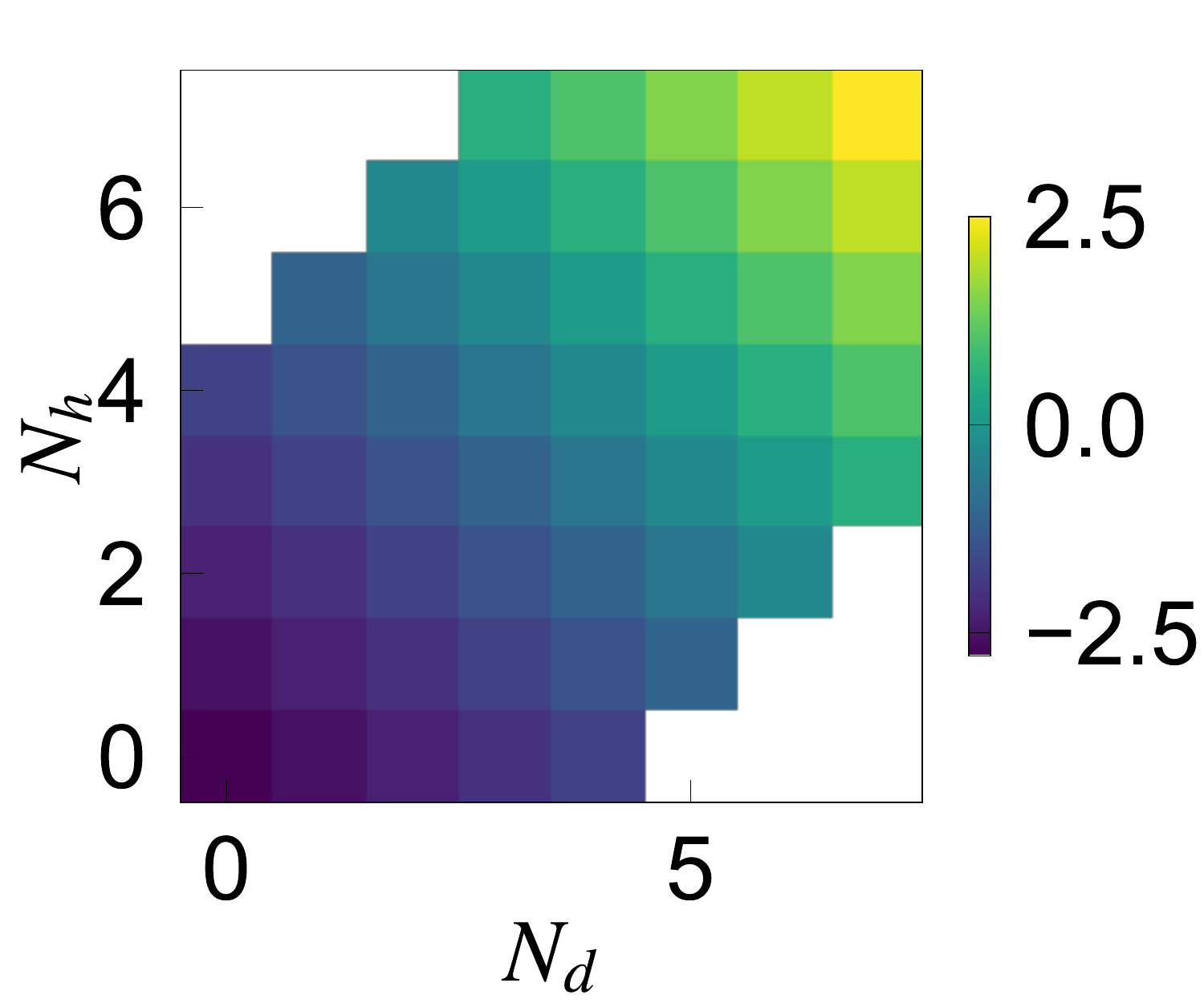} 
  \caption{Energy as a function of $N_d$ and $N_h$ for $\hat{H}_{\rm eff}/N$. Here, we use $U=10,J_{\rm ex}=0.4,V=0$ and $N=14$. }
  \label{fig:ED_energy_V0}
\end{figure}

\begin{figure}[t]
  \centering
    \hspace{-0.cm}
    \vspace{0.0cm}
\includegraphics[width=87mm]{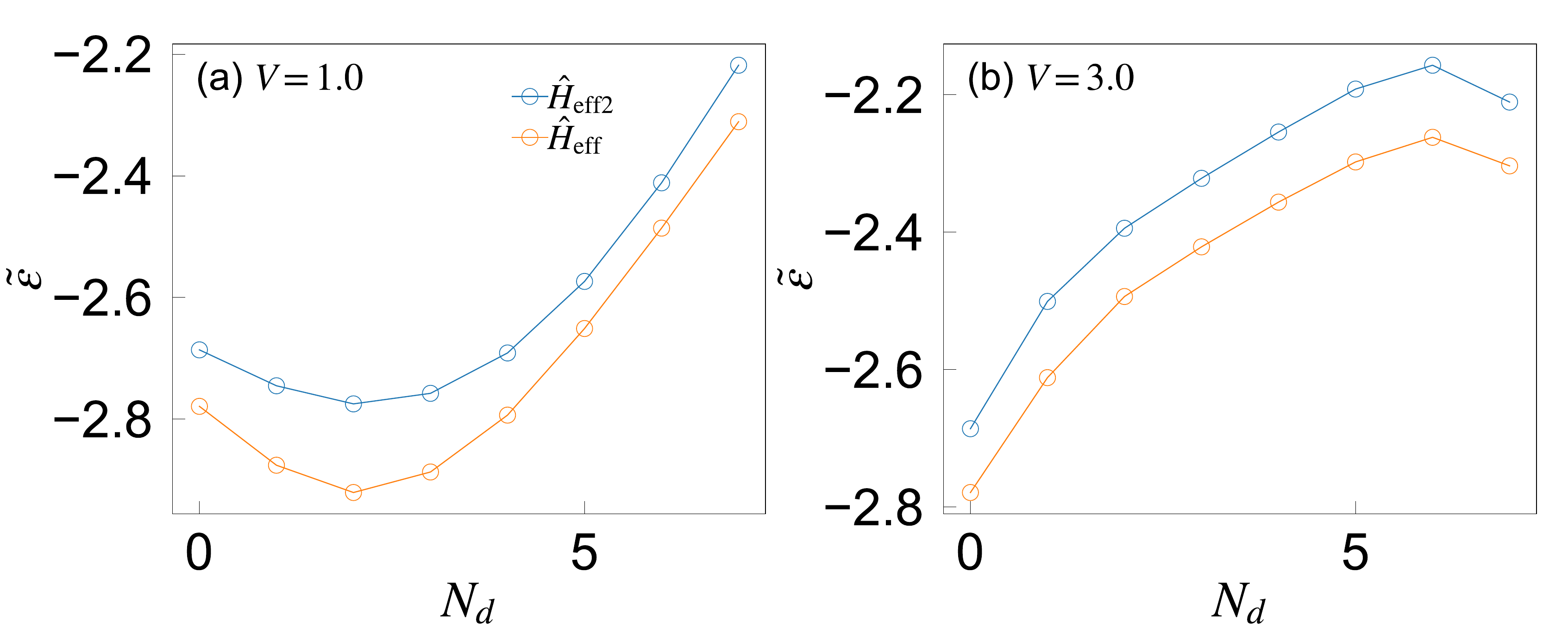} 
  \caption{Energy along $N_d=N_h$ as a function of $N_d$. Here, we use $U=10,J_{\rm ex}=0.4$ and $N=14$. To clarify the curvature of the function, we 
  plot $\tilde{\epsilon}\equiv \langle \hat{H}_{\rm eff} \rangle/N -(U-2V) N_d/N$ for $ \hat{H}_{\rm eff}$ and $\tilde{\epsilon}\equiv \langle \hat{H}_{\rm eff2} \rangle/N - (U-2V)N_d/N$ for $ \hat{H}_{\rm eff2}$.}
  \label{fig:ED_energy_V1_3}
\end{figure}

 In the normal equilibrium system, one may encounter a possible phase separation into a 
 high density and low density region. 
 In the present study of the photo-doped $U$-$V$ Hubbard model, there are two types of possible phase separations: (i) separation into a 
  high density and low density region,
 and (ii) separation into a doublon-holon rich region and a doublon-holon poor region.
 The stability against these phase separations can be discussed by expressing the energy (free energy for finite temperature) surface as a function of $N_d$ and $N_h$,
 and checking whether the surface is concave or not. 
 The analytical discussion in Sec.~\ref{sec:lemmas} implies that for $V=0$, the energy is flat along the line of $N_d+N_h={\rm constant}$ around the homogeneous solution with $\eta^2\neq 0$ at half-filling.
This means that the state is on the verge of phase separation into a 
high density and low density 
region.
 This is numerically confirmed in Fig.~\ref{fig:ED_energy_V0}. 
 We note that the energy surface is concave along  $N_d=N_h$ for small $V$, which means that the homogeneous solution is stable against the separation into a doublon-holon rich region and a doublon-holon poor region,
 see Fig.~\ref{fig:ED_energy_V1_3}(a).
 On the other hand, when $V$ is large, the function becomes convex, which is indicative of phase separation.
 We note that this unstable regime corresponds to the regime where the emergence of a doublon-holon cluster was previously reported \cite{Takahashi2005PRB}.

 \subsection{Correlation functions and response functions}
\begin{figure}[t]
  \centering
    \hspace{-0.cm}
    \vspace{0.0cm}
\includegraphics[width=87mm]{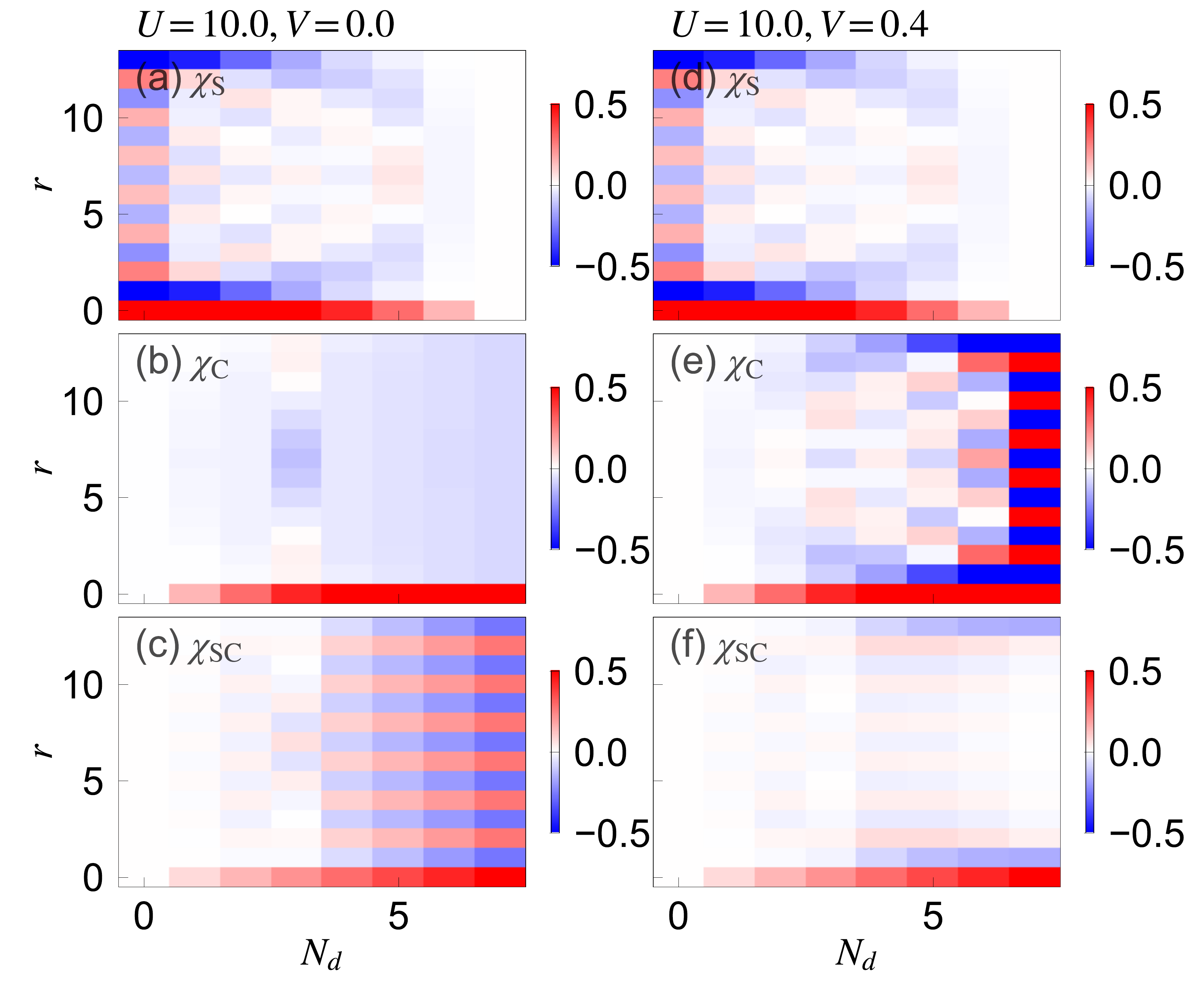} 
  \caption{Correlation functions for photo-doped states described by $\hH_{\rm eff}$ evaluated by ED for $U=10$ and $N=14$. 
 Panels (a)(d) show the charge correlation $\chi_{\rm c}(r)$, (b)(e) the spin correlation $\chi_{\rm s}(r)$ and (c)(f) the SC correlation $\chi_{\rm sc}(r)$. For (a-c), $V=0$ and for (d-f) $V=0.4$. }
  \label{fig:ED_full_corr}
\end{figure}

\begin{figure}[t]
  \centering
    \hspace{-0.cm}
    \vspace{0.0cm}
\includegraphics[width=87mm]{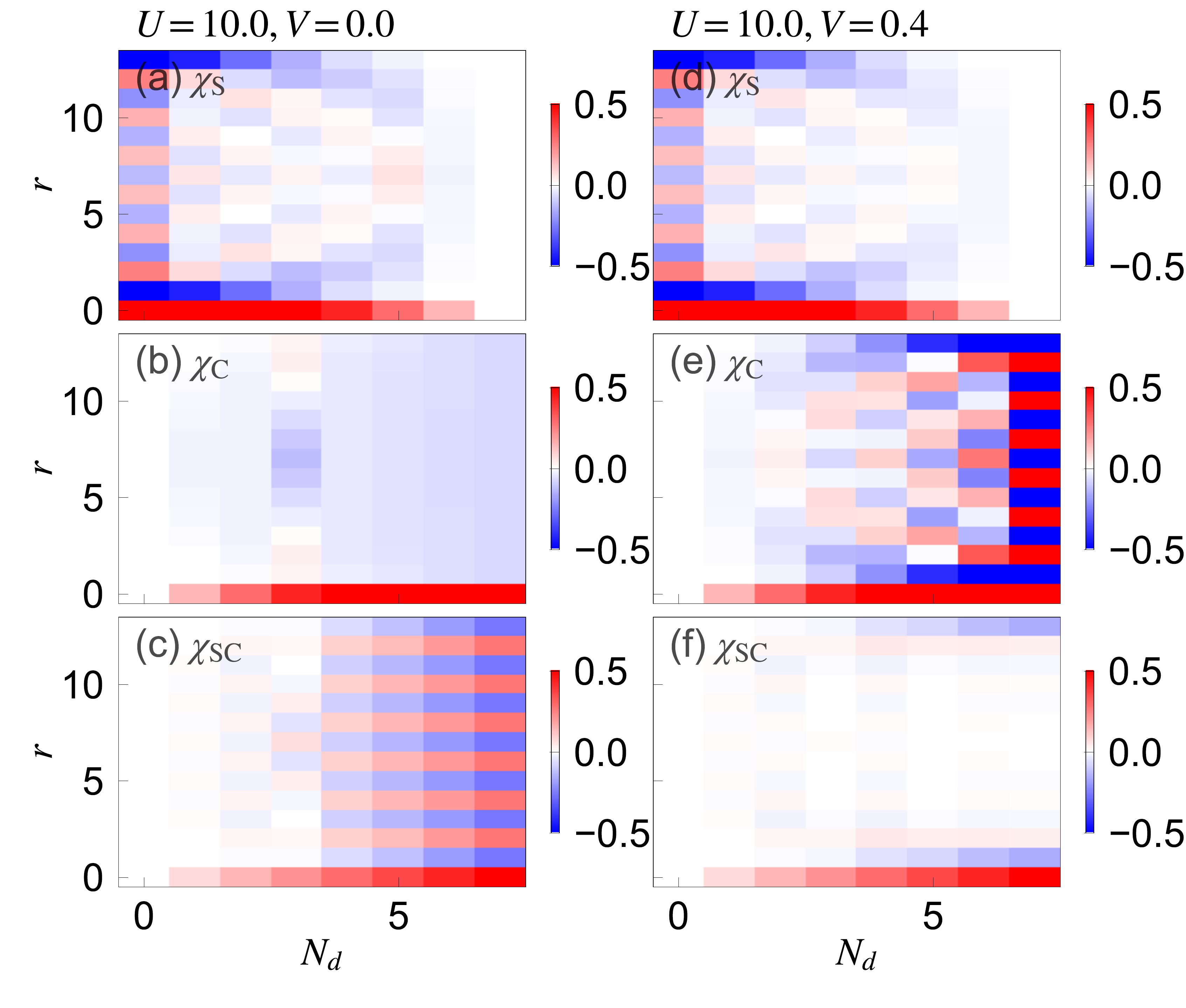} 
  \caption{Correlation functions for photo-doped states described by $\hH_{\rm eff2}$ evaluated by ED for $U=10$ and $N=14$. Panels (a)(d) show the charge correlation $\chi_{\rm c}(r)$, (b)(e) the spin correlation $\chi_{\rm s}(r)$ and (c)(f) the SC correlations $\chi_{\rm sc}(r)$. For (a-c), $V=0$ and for (d-f) $V=0.4$. }
  \label{fig:ED_simple_corr}
\end{figure}

In Figs.~\ref{fig:ED_full_corr} and ~\ref{fig:ED_simple_corr}, we show the correlation functions for the photo-doped states evaluated with ED for  $\hH_{\rm eff}$ and $\hH_{\rm eff2}$, respectively.
We focus here on half-filling.
First of all, one can see that the effect of $\hH_{\rm 3-site}$ is not essential since the spatial patterns obtained for $\hH_{\rm eff}$ and $\hH_{\rm eff2}$ match nicely.
Slight differences can be seen for the case of $V=0.4$, where the charge (SC) correlation is slightly stronger (weaker) for $\hH_{\rm eff2}$.

In both cases, for $V=0$, the system shows commensurate SDW correlations without photo-doping ($N_d=0$).
As we photo-dope the system, the spin correlations $\chi_{\rm s}$ become weaker and their spatial pattern deviates from the commensurate one.
While no clear pattern emerges for the charge correlations $\chi_{\rm c}$, a staggered pattern appears in the SC correlations $\chi_{\rm sc}$. 
The latter is the indication of $\eta$-pairing states, which are energetically favored due to $\hH_{\rm ex,dh}$.
The development of the SC pattern starts already for low photo-doping.
When $V$ is switched on, the behavior of the spin correlations is little affected. (Remember that $U\gg V$.)
On the other hand, some structure develops in $\chi_{\rm c}$ with increasing photo-doping and a commensurate CDW  is found in the extreme photo-doping regime ($N_d=N/2$).
There the model becomes the XXZ model.
The SC correlations are weakened by $V$, but the commensurate correlations do not vanish completely, in particular at  intermediate dopings.
These results are all consistent with the iTEBD simulations for the thermodynamic limit.

\begin{figure}[t]
  \centering
    \hspace{-0.cm}
    \vspace{0.0cm}
\includegraphics[width=87mm]{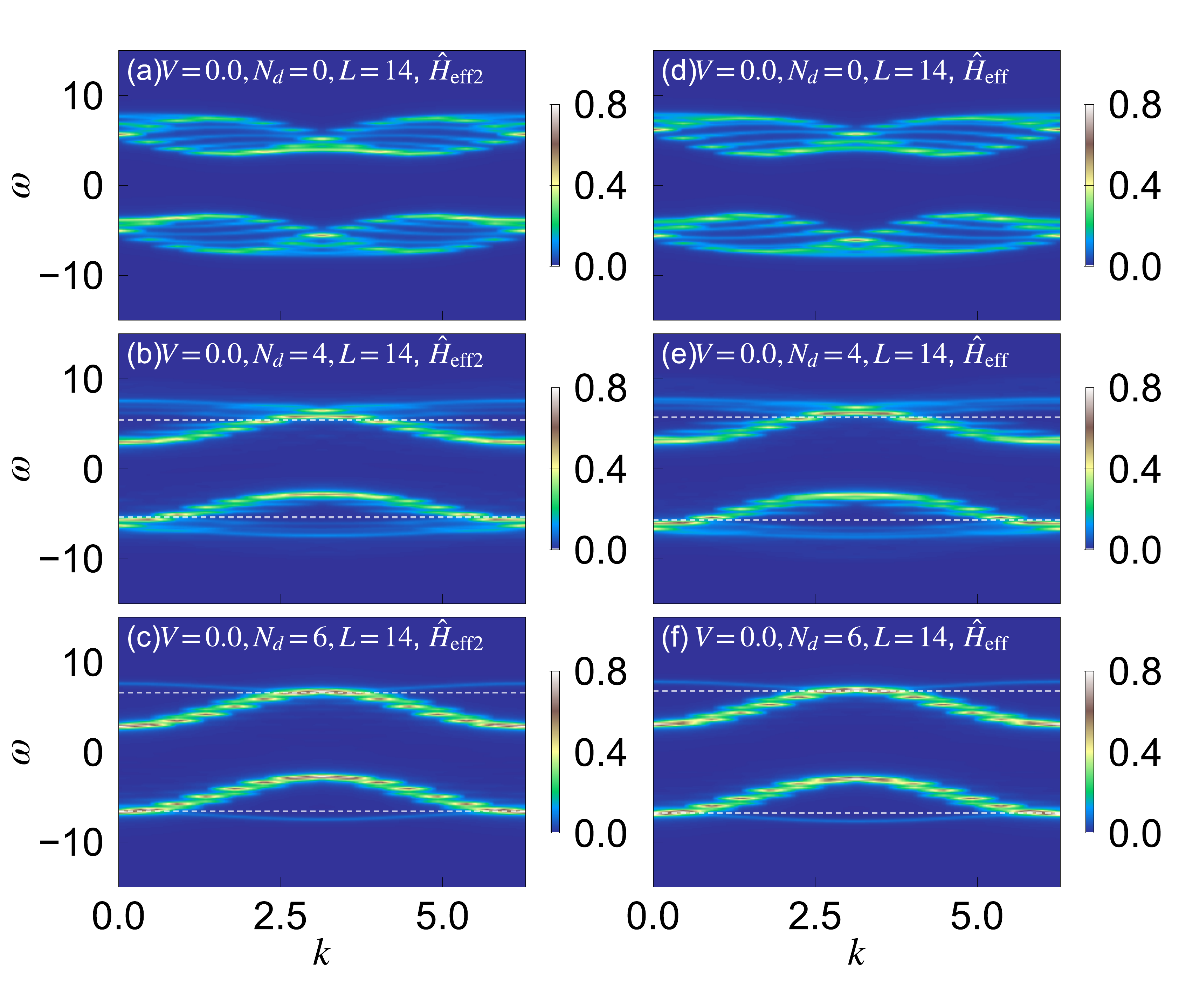} 
  \caption{Momentum-resolved single-particle spectra $A_k(\omega)$ for the indicated parameters and models evaluated by ED. Here, we use $U=10,V=0.0, J_{\rm ex}=0.4$. The dashed lines show the Fermi levels for the UHB and LHB.}
  \label{fig:ED_Akw_V0}
\end{figure}

\begin{figure}[t]
  \centering
    \hspace{-0.cm}
    \vspace{0.0cm}
\includegraphics[width=87mm]{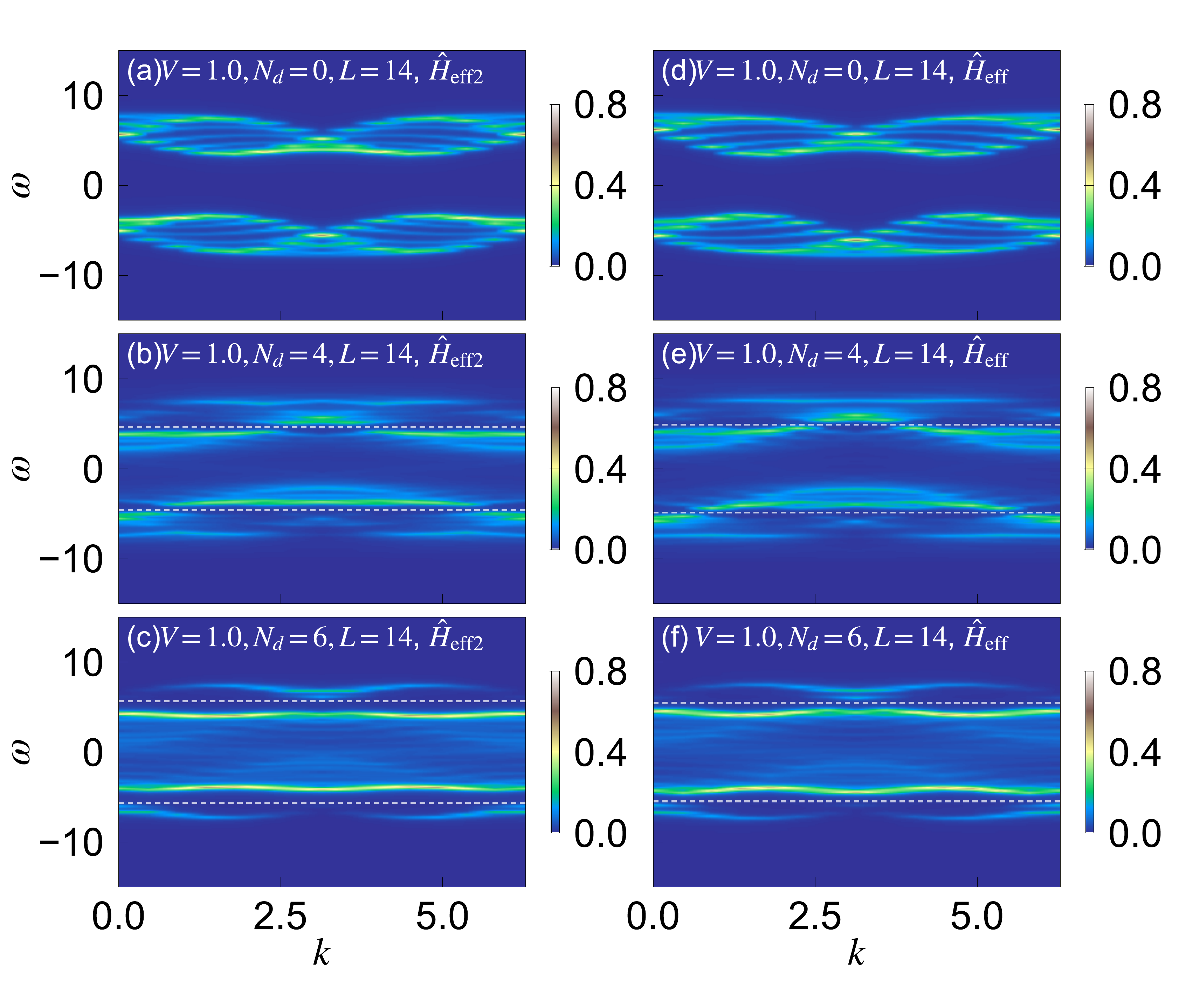} 
   \caption{Momentum-resolved single-particle spectra $A_k(\omega)$ for indicated parameters and models evaluated by ED. Here, we use $U=10,V=1.0, J_{\rm ex}=0.4$. The dashed lines show the Fermi levels for the UHB and LHB.}
  \label{fig:ED_Akw_V1}
\end{figure}

One can furthermore see that the effects of $\hH_{\rm 3-site}$ is also small for dynamical properties such as single-particle spectra.
In Fig.~\ref{fig:ED_Akw_V0} and Fig.~\ref{fig:ED_Akw_V1}, we compare the momentum-resolved spectral functions ($A_{k}(\omega)$) for different photo-dopings and interactions for $\hH_{\rm eff}$ and $\hH_{\rm eff2}$.
The characteristic features in the spectral functions match well between the two models.
Although it is difficult to tell because of the finite system size, one cannot identify a clear gap at the ``Fermi levels" of the upper Hubbard band (UHB) and the lower Hubbard band (LHB)
for $V=0$, i.e. $\eta$-pairing states, see Fig.~\ref{fig:ED_Akw_V0}.
It is also possible to identify a flattening of the band structure just above (below) the Fermi levels in the UHB (LHB), consistent with the peak structure observed in the iTEBD spectra around the Fermi levels.
Furthermore, one can see the opening of gaps at the Fermi levels for the UHB and LHB for $V=1.0$, where (incommensurate) CDW correlations develop, see Figs.~\ref{fig:ED_Akw_V1} (b)(c)(e)(f).
These features are all consistent with the iTEBD analysis.

 \begin{figure}[t]
  \centering
    \hspace{-0.cm}
    \vspace{0.0cm}
\includegraphics[width=87mm]{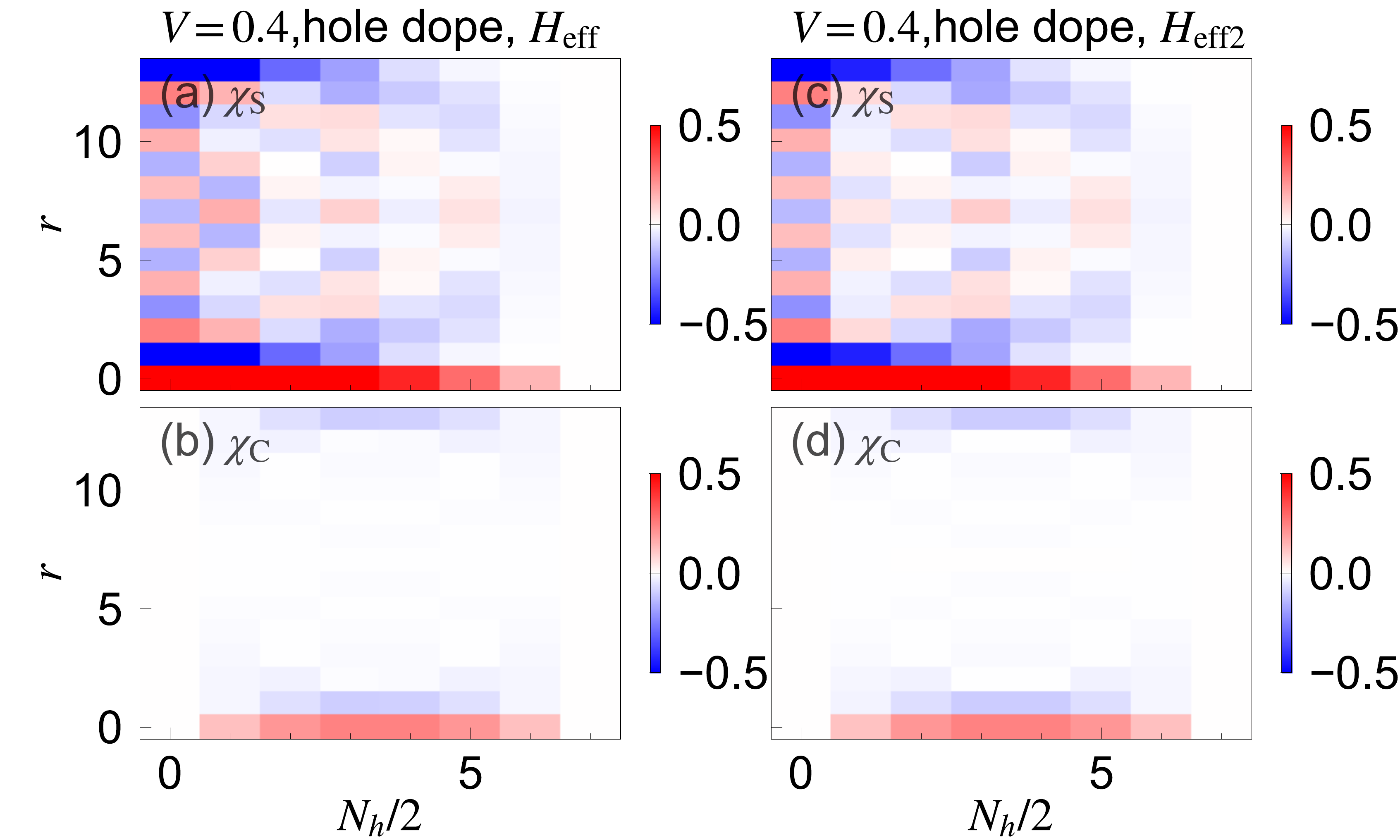} 
  \caption{Correlation functions for hole-doped states evaluated by ED for $U=10.0,V=0.4$ and $N=14$. Panels (a)(c) show  the spin correlation functions $\chi_{\rm s}(r)$, and panels (b)(d) the charge correlation functions $\chi_{\rm c}(r)$. (a)(b) are for the states described by $\hH_{\rm eff}$, while (c)(d) are for those described by $\hH_{\rm eff2}$.}
  \label{fig:ED_simple_full_comp_holedope_V04}
\end{figure}

 \begin{figure}[t]
  \centering
    \hspace{-0.cm}
    \vspace{0.0cm}
\includegraphics[width=87mm]{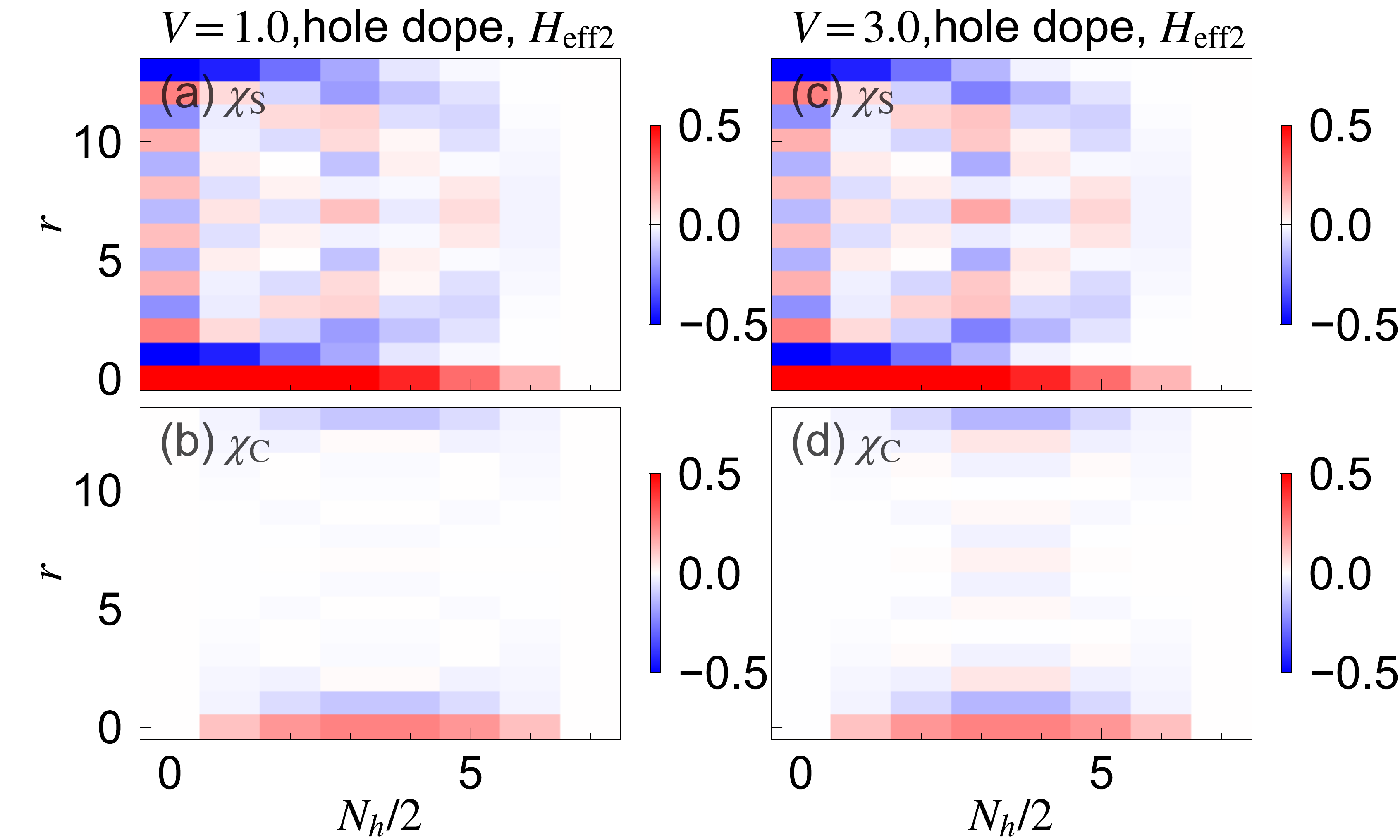} 
  \caption{Correlation functions for hole-doped states described by $\hH_{\rm eff2}$. Here we use ED for  $U=10.0$ and $N=14$. Panels (a)(c) show the spin correlation functions $\chi_{\rm s}(r)$, and (b)(d) the charge correlation functions $\chi_{\rm c}(r)$. (a)(b) are for $V=1.0$, while (c)(d) are for $V=3.0$.}
  \label{fig:hole_dope_Vdep}
\end{figure}

 \begin{figure}[t]
  \centering
    \hspace{-0.cm}
    \vspace{0.0cm}
\includegraphics[width=87mm]{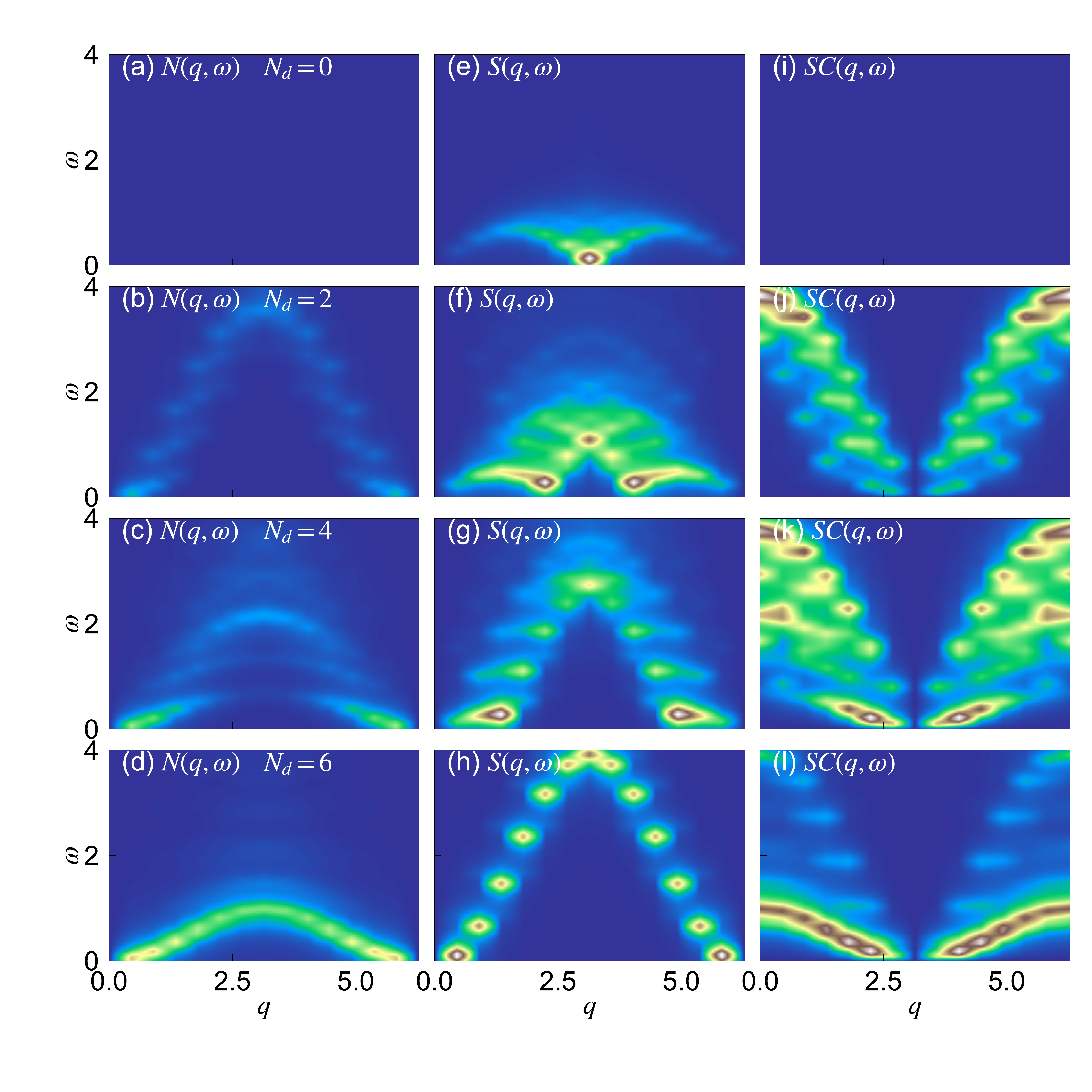} 
  \caption{Response functions of photo-doped states described by $\hH_{\rm eff2}$ for specified doping levels. Here, we use ED and $U=10,V=0$ and $N=14$. Panels (a-d) show the charge response function $N(q,\omega)$,
  (e-h)  the spin response function $S(q,\omega)$, and (i-l) the SC response function $SC(q,\omega)$.}
  \label{fig:RF_V0}
\end{figure}

We next show additional results comparing photo-doped and chemically doped systems.
In Fig.~\ref{fig:ED_simple_full_comp_holedope_V04}, we show the results of $\hH_{\rm eff}$ and $\hH_{\rm eff2}$ for hole-doped cases with $V=0.4$.
Again one can see that the effects of the three site term are small.
In Fig.~\ref{fig:hole_dope_Vdep}, we show the results for different values of $V$.
While the pattern of the spin correlations remains almost the same, one can see that the intensity of the charge correlation is slightly increased with increasing $V$.

Finally, we discuss the behavior of the (linear) response functions~\cite{Dagotto1994RMP}:
\eqq{
N(q,\omega) &=  -i\int_0^\infty dt e^{i\omega t} e^{-\gamma t} \langle [\hn_q(t)^\dagger, \hn_q(0)] \rangle ,\\
S(q,\omega) &= -i\int_0^\infty dt e^{i\omega t} e^{-\gamma t} \langle [\hS^z_q(t)^\dagger, \hS^z_q(0)] \rangle , \\
SC(q,\omega) &= -i\int_0^\infty dt e^{i\omega t} e^{-\gamma t} \langle [\hat{\Delta}_q(t)^\dagger, \hat{\Delta}_q(0)] \rangle .
}
Here, $\hS^z_q = \frac{1}{\sqrt{N}} \sum_j e^{-iq j} \hs^z_j$, $\hn_q = \frac{1}{\sqrt{N}} \sum_j e^{-iq j} \hn_j$, and $\hat{\Delta}_q = \frac{1}{\sqrt{N}} \sum_j e^{-iq j} \hat{\Delta}_j$.
$\gamma$ is a damping factor which we set to $0.15$.
Since one can confirm again that the effect of the three site term is minor for these quantities, we only show results for $\hH_{\rm eff2}$.
Figure~\ref{fig:RF_V0} presents the response functions for $V=0$.
The charge response function $N(q,\omega)$ always shows a gapless mode at $q=0$.
In particular, for large doping, the energy scale of the dispersion of the collective mode becomes $J_{\rm ex}$.
The spin response function $S(q,\omega)$ shows a massless mode at $q=\pi$ without photo-doping, which is consistent with the development of quasi-long range commensurate SDW correlation.
Also, the signals in $S(q,\omega)$ represent the spinon dispersion, whose energy scale is $J_{\rm ex}$ for the undoped case.
For large doping the energy scales become $v$ and the signature resembles the free particle dispersion, see Fig.~\ref{fig:RF_V0}(h).
In between, one can identify structures where the above two features are mixed.
These presumably originate from the dynamics of the kink of the spin configuration and the kinetics of the particles that hold the spin (singlon).
As we increase the doping, the massless mode shifts to some incommensurate value of $q$.
As for the SC response function $SC(q,\omega)$, one can identify a gapless mode emerging from $q=\pi$ and the intensity profile looks similar to that of $N(q,\omega)$ shifted by $\pi$.

 \begin{figure}[t]
  \centering
    \hspace{-0.cm}
    \vspace{0.0cm}
\includegraphics[width=87mm]{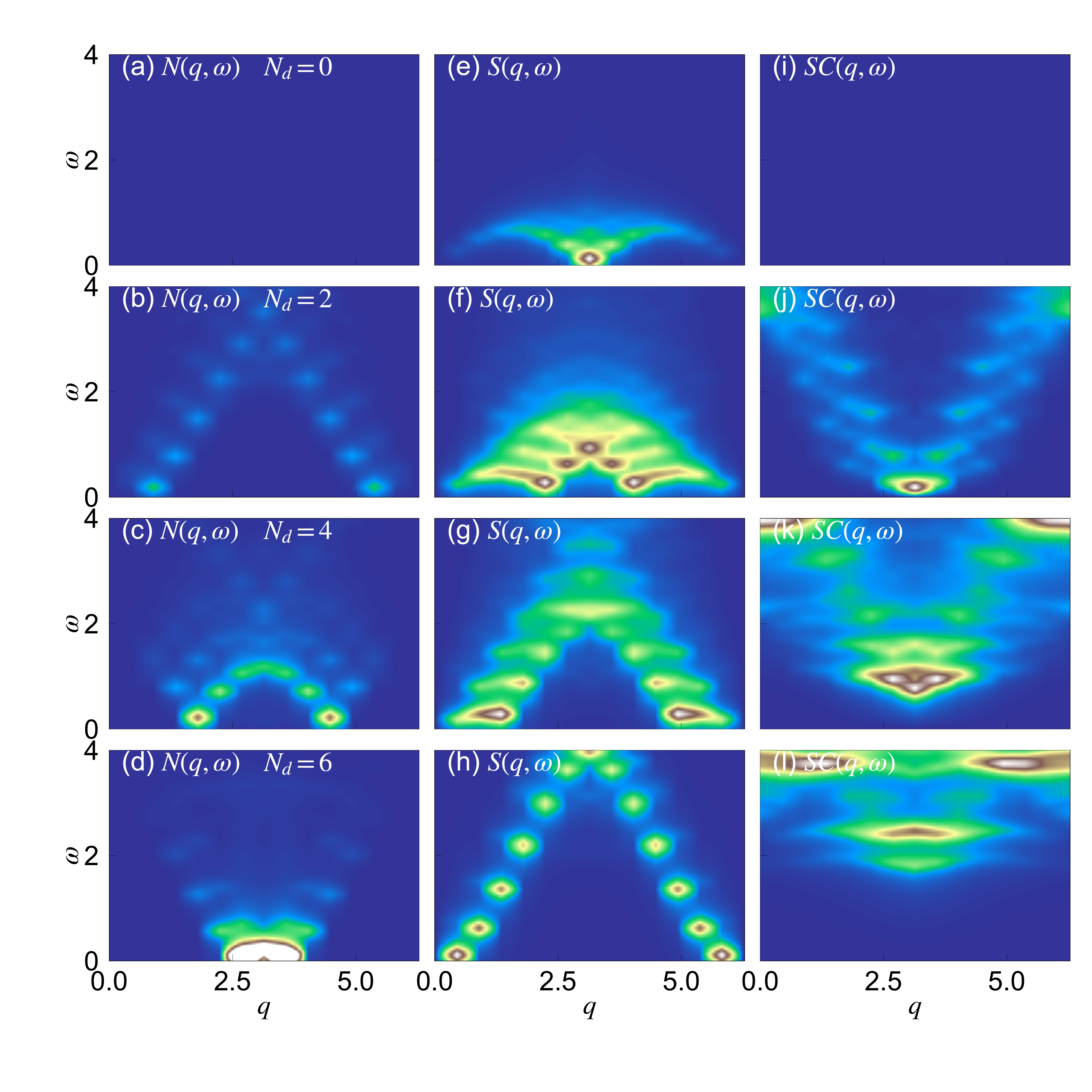} 
  \caption{Response functions of photo-doped states described by $\hH_{\rm eff2}$ for specified doping levels. Here, we use ED and $U=10,V=1$ and $N=14$. Panels (a-d) show the charge response function $N(q,\omega)$,
  (e-h) the spin response function $S(q,\omega)$, and (i-l) the SC response function $SC(q,\omega)$.}
  \label{fig:RF_V1}
\end{figure}

\begin{figure}[h]
  \centering
    \hspace{-0.cm}
    \vspace{0.0cm}
\includegraphics[width=87mm]{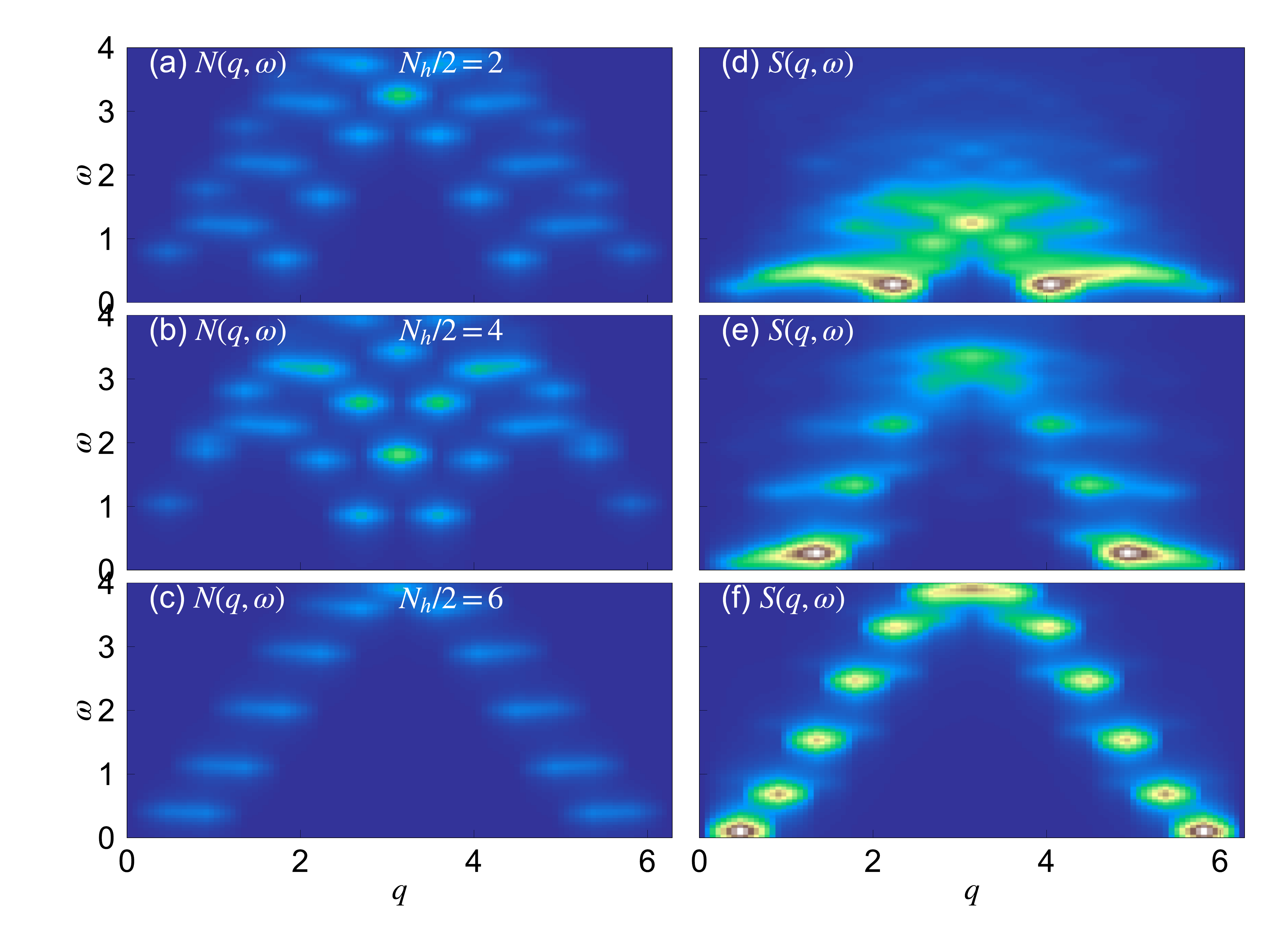} 
  \caption{Response functions of hole-doped states described by $\hH_{\rm eff2}$ for the specified doping levels. Here, we use ED, $U=10$, $V=1$ and $N=14$. Panels (a-c) show the charge response function $N(q,\omega)$, and
  (d-f)  the spin response function $S(q,\omega)$.}
  \label{fig:RF_V1_hole}
\end{figure}

In Fig.~\ref{fig:RF_V1}, we show the results for $V=1$.
The behavior of $S(q,\omega)$  is similar to that for $V=0$.
On the other hand, $N(q,\omega)$ shows massless modes at incommensurate values of $q$,
which is consistent with  the development of quasi-long range incommensurate CDW correlations.
The SC response function shows a clear nonzero gap, which is consistent with the exponential decay of the SC correlations. 
In Fig.~\ref{fig:RF_V1_hole}, we show the results for $V=1$ for the hole-doped system.
The behavior of $S(q,\omega)$  is similar to that for photo-doping.
As for $N(q,\omega)$, there is no clear development of low-lying modes, which is consistent with weak CDW signals in the correlation function.

\end{document}